# Effect of Triangular Pre-Cracks on the Mechanical Behavior of 2D MoTe$_2$: A Molecular Dynamics Study

Md. Jobayer Aziz[1], Md Akibul Islam[2*], Md. Rezwanul Karim[1], Arafat Ahmed Bhuiyan[1]

[1]Department of Mechanical and Production Engineering, Islamic University of Technology (IUT), Board Bazar, Gazipur 1704, Bangladesh

[2]Department of Mechanical and Industrial Engineering, University of Toronto, Canada

*Correspondence: aislam@mie.utoronto.ca (M.A.I.)

## Abstract

Among two-dimensional (2D) materials, transition metal dichalcogenides (TMDs) stand out for their remarkable electronic, optical, and chemical properties. In addition to being variable bandgap semiconductor materials, the atomic thinness provides flexibility to TMDs. Therefore, understanding the physical properties of TMDs for applications in flexible and wearable devices is crucial. Despite the growing enthusiasm surrounding two-dimensional transition metal dichalcogenides (TMDs), our understanding of the mechanical characteristics of molybdenum ditelluride (MoTe$_2$) remains limited. The mechanical properties of MoTe$_2$ deteriorate in the presence of pre-existing cracks or vacancy defects, which are very common in grown TMDs. In this study, the fracture properties and crack propagation of monolayer molybdenum ditelluride (MoTe$_2$) sheets containing pre-existing triangular cracks with various vertex angles are investigated by performing molecular dynamics (MD) simulations of uniaxial and biaxial tensile loading. Due to pre-crack length, angle, and perimeter variations, monolayer MoTe$_2$ with pre-existing cracks underwent considerable changes in Young's modulus, tensile strength, fracture toughness, and fracture strain values. We have found that the pre-cracked MoTe$_2$ is more brittle than its pristine counterpart. Regulated alteration of pre-crack angle under constant simulation conditions improved the uniaxial mechanical properties. Similarly, regulated alteration of the perimeter of the pre-crack resulted in improved biaxial mechanical properties. This study contributes to the foundational knowledge for advanced design strategies involving strain engineering in MoTe$_2$ and other similar transition metal dichalcogenides.

## 1. Introduction

The discovery of graphene has triggered significant interest in the study of two-dimensional (2D) materials [1,2]. While graphene possesses remarkable mechanical [3,4], magnetic [5], electronic [6], optical [7], thermal [8], and tribological [9,10] properties, its lack of bandgap [11] renders it unsuitable for applications in field-effect transistors (FETs) and nanoelectronics. Another group of atomically thin materials, transition metal dichalcogenides (TMDs), have garnered attention for their semiconducting properties and direct bandgap [12–16], which makes them attractive for nanoelectronics applications. TMDs consist of transition metal atoms (e.g., M =, W, Ni) sandwiched between two chalcogen atom (e.g., X= S, Se, Te) sub-layers, resulting in MX$_2$ atomic structure. Prominent examples of TMDs include MoS$_2$, MoTe$_2$, MoSe$_2$, WS$_2$, WTe$_2$, and WSe$_2$. Among these TMDs, MoTe$_2$ stands out with its relatively narrow bandgap of ~1.1 eV [17], making it suitable for the development of 2D devices and heterostructures [18–21]. Like many TMDs, it can exhibit different phases: the thermodynamically favorable semiconducting 2H phase [22], the metastable metallic high symmetry 1T phase [23,24], and the distorted low symmetry 1T' phase [22]. Experimental findings indicate that the 2H phase of $MoTe_2$ has the highest stability under ambient conditions [23,25,26]. Therefore, our focus lies on studying $MoTe_2$ in its 2H phase.

$MoTe_2$ can be synthesized using top-down methods like mechanical and liquid exfoliation [27–32], as well as bottom-up methods such as molecular beam epitaxy (MBE) [33,34], chemical vapor deposition (CVD) [35,36], chemical vapor transport [37,38], physical vapor deposition (PVD) [39], electrodeposition [40,41], thermal treatment [42–45], flux [46] and Tellurization of Molybdenum or Molybdenum Oxide [36,47]. Both naturally occurring and fabricated $MoTe_2$ samples contain different types of defects [33,48–57], which are introduced during industrial processes or radiation damage [58,59]. Pre-existing cracks and defects play crucial role in shaping the mechanical, electronic, optical, and magnetic behaviors of $MoTe_2$. This underscores the importance of conducting thorough investigations into the mechanical properties of $MoTe_2$ containing pre-existing cracks and defects.

$MoTe_2$ is suitable for applications in flexible and stretchable electronics due to its excellent electronic, mechanical, and optical properties [60–63]. In TMDs electronic states are controlled by externally applied uniaxial and biaxial strain [64,65]. Hence, Stretchable electronic devices are designed according to constituent materials' uniaxial and biaxial tensile properties [66]. Both experimental and computational methods have been employed in recent years to investigate the mechanical properties of $MoTe_2$. Sun *et al.* [67] employed nanoindentation experiments and first-principles calculations to investigate the elastic properties and fracture behavior of biaxially deformed $MoTe_2$, revealing reduced breaking strengths in distorted phases due to uneven bonding distribution and resulting distinct fracture patterns. Mortazavi *et al.* [68] explored the mechanical properties of single-layer $MoTe_2$ structures, uncovering non-isotropic mechanical responses, tunability of electronic properties under mechanical strain in 2H-MoTe$_2$, the auxetic behavior of 1T-MoTe$_2$, and the high strength of 1T'/2H-MoTe$_2$ heterostructures. Pristine and defective TMDs are often subjected to uniaxial and biaxial tensile loading to investigate their mechanical properties [68–74]. Pereira Júnior et al. [75] investigated the uniaxial tensile properties of 2H and 1T phases of MoTe$_2$, MoS$_2$, and MoSe$_2$ using reactive molecular dynamics simulations. Jiang and Zhou [76] developed a Stillinger Weber interatomic potential to investigate the mechanical properties of 2H and 1T-MoTe$_2$. According to their findings, 2H-MoTe$_2$ has Young's modulus of 79.8 N/m and 78.5 N/m along the armchair and zigzag directions, respectively. For 1T-MoTe$_2$, Young's modulus is 81.6 N/m and 81.2 N/m along the armchair and the zigzag directions, respectively. This result closely resembles experimental results of 79.4 N/m from Ref. [77], or 87.0 N/m from Ref. [78] for 2H-MoTe$_2$, and ab initio result of 92 N/m [79] for 1T-MoTe$_2$.

Researchers have found that the mechanical properties of TMDs degrade in the presence of pores and pre-cracks [70]. Pre-existing cracks are generally triangle-shaped [80–89]. However, the effect of such cracks on single-layer 2H-MoTe$_2$ has not been well studied. This study uses molecular dynamics software, LAMMPS [90], to investigate MoTe$_2$ with triangle-shaped pre-cracks under different tensile loading conditions. Specifically, we have determined the stress-strain relationship through uniaxial and biaxial tensile loading and derived the mechanical properties resulting from uniaxial and biaxial tensile loading at room temperature. Additionally, we have provided insights into the fracture mechanism through visualization. By examining the mechanical properties of pristine MoTe$_2$ and MoTe$_2$ containing triangular pre-cracks, we have developed a comprehensive understanding of the fracture behavior of monolayer MoTe$_2$ sheets with triangular vacancy defects.

## 2. Methodology

Unlike single-layer graphene, MoTe$_2$ is composed of three atomic sub-layers. One single layer consists of two Te sub-layers flanking one Mo sub-layer in the *Te–Mo–Te* order (shown in **Figure 1**). In 2H-MoTe$_2$, Te atoms of a sub-layer are directly situated above or below the Mo atoms of the adjacent layer.

In 2H-MoTe$_2$, Mo and Te atoms are arranged in trigonal prismatic coordination, with each Mo atom encircled by six neighboring Te atoms, and each Te atom establishing bonds directly with three Mo atoms, within each layer [91]. These intralayer atoms establish strong covalent bonds, while the interlayer atoms are connected through weaker van der Waals interactions. The lattice constant of MoTe$_2$ was 3.55 Å. Mo-Te bond length was 2.73 Å [68,92].

Three bond-stretching (Mo–Te, Mo–Mo, and Te–Te) and three angle-bending motions (two Mo–Te–Te angles with Mo as the central atom and one Te–Mo–Mo angle) are found in the covalent bond. Using MATLAB [93] script, we generated 27 nm × 27 nm MoTe$_2$ sheets from an orthogonal unit cell (with b=√3a and α = β = γ = 90°). The nanosheet had an effective thickness of approximately 0.6 nm. The initial system contained 13376 Te atoms and 6688 Mo atoms, a total 20064 atoms, before introducing the pre-cracks manually with the help of Python programming language [94].

Pre-cracks in the monolayer MoTe$_2$ were created through the selective removal of atoms, forming triangular vacancy defects resembling isosceles triangles. Similar to a prior investigation [95], only

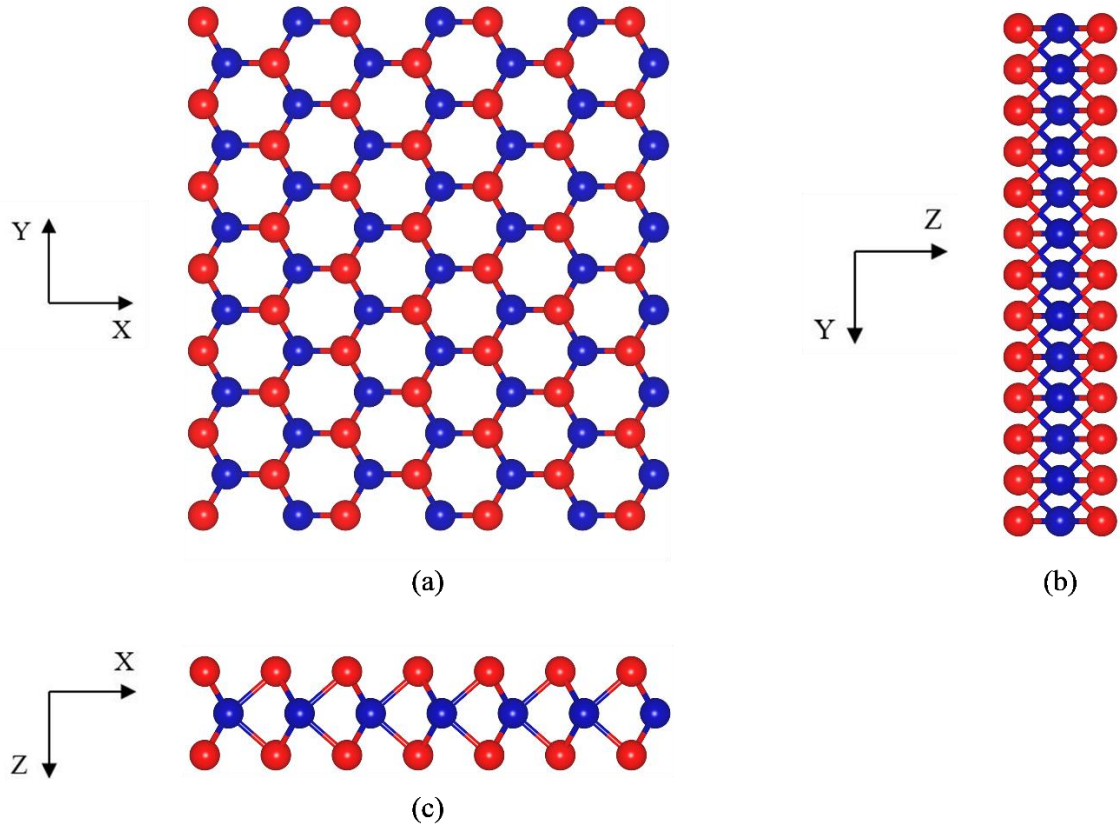

*Figure 1* *(a) Top view, (b) side view and (c) front view of the atomic structure of a single layer MoTe$_2$. The blue atoms represent Mo, and the red atoms represent Te. The X-axis points in the direction of the armchair orientation, while the Y-axis points in the direction of the zigzag orientation.*

one pre-crack was introduced at the center of each monolayer MoTe$_2$ sheet. The height and the base of the isosceles triangles are denoted by 'H' and 'W' respectively ( **Figure *2*** (a), (d)). The base 'W' of the isosceles triangle is located along the X-axis (armchair axis) ( **Figure *2*** (a-c)) or the Y-axis (zigzag axis) ( **Figure *2*** (d-f)) of the MoTe$_2$. Other than the base, the remaining two sides of that isosceles triangle are equal in length. In the isosceles triangle, the angle opposite to the triangle's base 'W' is referred to as the vertex angle. This vertex angle 'θ' is measured in degree

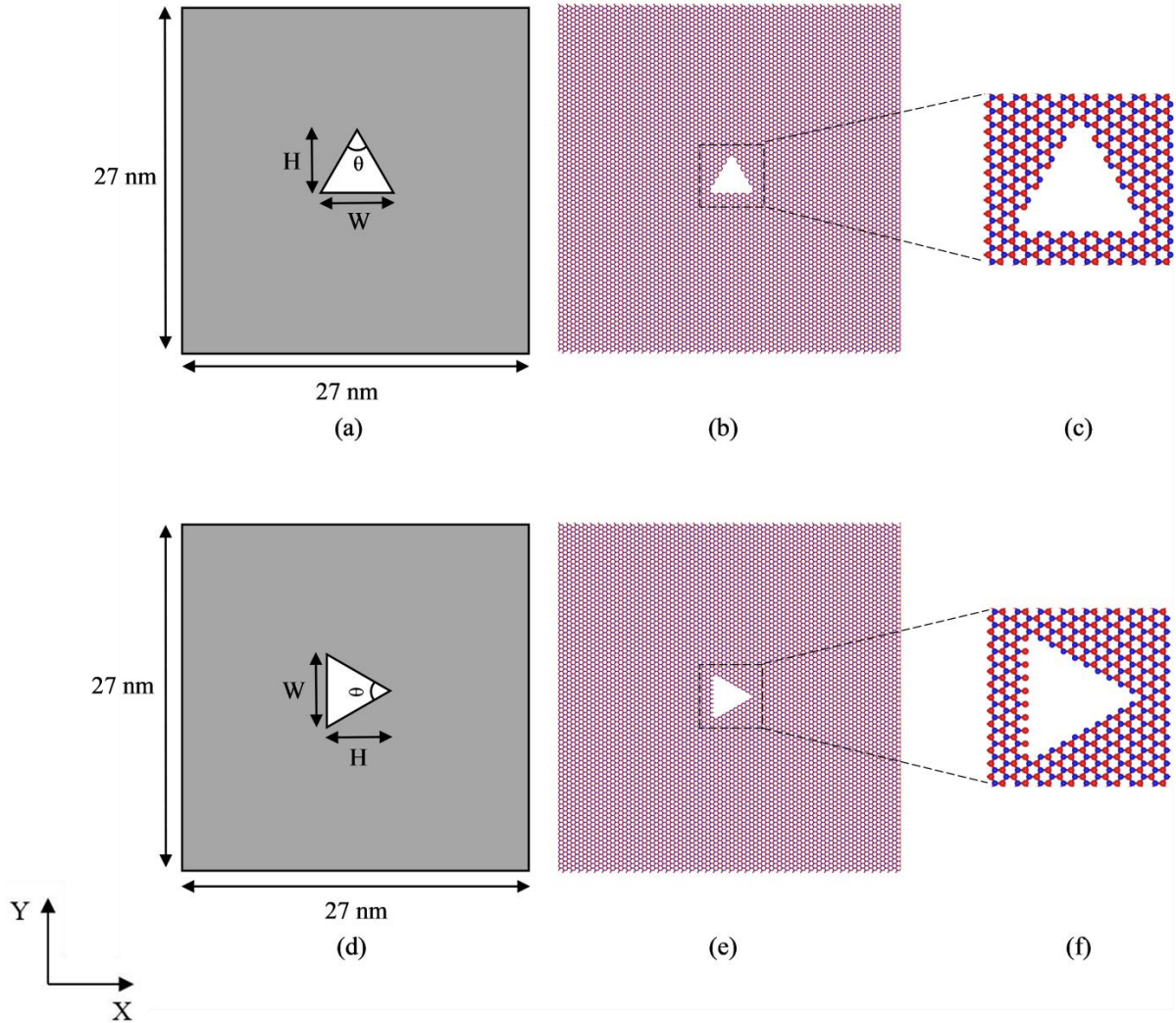

***Figure 2*** *(a) Illustration of the system size and configuration of the X60° system, which is a single layer MoTe$_2$ of dimensions 27 nm × 27 nm along the X-Y axis with an isosceles-shaped pre-crack at its center that has a vertex angle of θ = 60°. The base 'W' of this pre-crack triangle is aligned with the X-axis (armchair orientation), while pre-crack height or pre-crack length 'H' is aligned with the Y-axis (zigzag orientation). (b) Atomic structure of X60° system. (c) A zoomed-in view of the pre-crack in the X60° system. (d) Illustration of the system size and configuration of the Y60° system, which is a single layer MoTe$_2$ of dimensions 27nm × 27nm along the X-Y axis with an isosceles-shaped pre-crack at its center that has a vertex angle of θ = 60°. The base 'W' of this*

*pre-crack triangle is aligned with the Y-axis (zigzag orientation), while pre-crack height or pre-crack length 'H' is aligned with the X-axis (armchair orientation). (e) Atomic structure of Y60° system. (f) A zoomed-in view of the pre-crack in the Y60° system.*

(°) unit. The height 'H' of the isosceles triangle is perpendicular to the base 'W'. Height 'H' can also be termed as the pre-crack length of the triangle-shaped pre-crack. As a convenient reference, we labelled the pre-cracked MoTe$_2$ containing isosceles triangle-shaped pre-crack where the base 'W' of the triangle is positioned along the X-axis and height 'H' along the Y-axis with vertex angle θ = 90° as "X90°" system. Since all the systems' porosity is approximately the same (**Table 1**), pre-crack length decreases when the vertex angle increases (see supplementary information **Figure S 1**, **Figure S 2**). During the uniaxial tensile loading, strain is applied along the same direction of the base 'W' of the pre-crack triangle. We investigated a total of fourteen systems containing isosceles triangle shaped pre-cracks with different vertex angles: X30°, X60°, X75°, X90°, X105°, X120°, X150° systems ("X_°" systems of supplementary information **Figure S 1**), where 'W' is parallel to X-axis (armchair orientation) and 'H' is parallel to Y-axis (zigzag orientation); and Y30°, Y60°, Y75°, Y90°, Y105°, Y120°, Y150° systems ("Y_°" systems of supplementary information **Figure S 2**), where 'W' is parallel to Y-axis (zigzag orientation) and 'H' is parallel to X-axis (armchair orientation). **Table 1** comprehensively summarizes the quantity of removed atoms and the corresponding defect percentages for all fourteen defective configurations.

*Table 1* The configuration of all fourteen MoTe$_2$ systems with triangular pre-existing cracks used in this study. Systems with base 'W' (of the triangular pre-crack) lying along the X-axis are denoted by "X_°" systems whereas systems with base 'W' (of the triangular pre-crack) lying along the Y-axis are denoted by "Y_°" systems. The perimeter of the pre-existing cracks were calculated using ImageJ [96]. For ease of comparison, systems with almost the same percentage of defect or porosity were chosen for the study.

| System Category | System Name | Vertex Angle | Atoms Removed | | | Perimeter (nm) | Porosity (%) |
|---|---|---|---|---|---|---|---|
| | | | Te | Mo | Total | | |
| X_° | X30° | 30° | 92 | 45 | 137 | 9.41 | 0.683 |
| | X60° | 60° | 90 | 45 | 135 | 8.64 | 0.673 |
| | X75° | 75° | 92 | 43 | 135 | 8.46 | 0.673 |
| | X90° | 90° | 90 | 46 | 136 | 9.09 | 0.678 |
| | X105° | 105° | 94 | 41 | 135 | 8.91 | 0.673 |
| | X120° | 120° | 90 | 45 | 135 | 9.78 | 0.673 |
| | X150° | 150° | 91 | 44 | 135 | 12.32 | 0.673 |
| Y_° | Y30° | 30° | 82 | 51 | 133 | 9.92 | 0.663 |
| | Y60° | 60° | 90 | 45 | 135 | 8.36 | 0.673 |
| | Y75° | 75° | 90 | 44 | 134 | 8.84 | 0.668 |
| | Y90° | 90° | 92 | 45 | 137 | 9.06 | 0.683 |
| | Y105° | 105° | 90 | 43 | 133 | 9.49 | 0.663 |
| | Y120° | 120° | 87 | 49 | 136 | 10.314 | 0.678 |
| | Y150° | 150° | 84 | 50 | 134 | 12.496 | 0.668 |

## 3. Computational Details

In the present study, molecular dynamics (MD) simulation technique was implemented to investigate the response of MoTe$_2$ under uniaxial and biaxial tensile loading. The simulations were carried out on Large-scale Atomic/Molecular Massively Parallel Simulator (LAMMPS) [90]. Afterwards, the fracture was visualized on OVITO [97].

To accurately model the mechanical behavior of 2D monolayer TMDs, the Stillinger-Weber (SW) interatomic potential developed by Jiang *et al.* [76] was selected. This potential is widely employed to predict mechanical responses of TMDs [70,75,98]. One notable advantage of the Stillinger-Weber (SW) potentials, compared to other potentials, is their ability to incorporate atomic interactions that encompass both two-body bond-stretching and three-body bond-bending interactions.

Periodic boundary conditions (PBC) were applied in all dimensions to mitigate the influence of free edges, ensuring that the results were independent of sample sizes. Cross-layer interactions were avoided in the out-of-plane direction by introducing 5 nm vacuum gaps above and below the lattice structure.

The simulation process involved initially minimizing equilibrium energy of the system using the conjugate gradient (CG) algorithm [99], followed by equilibration to a thermal steady state employing NPT (isothermal-isobaric) ensemble for 500 ps at room temperature (300K). A timestep of 1 fs was adopted for the whole simulation. For efficient equilibration both temperature and pressure damping constants were set to 20 ps. Subsequently, using NPT ensemble (constant number of atoms, pressure and temperature), tension was applied at a constant strain rate of $10^8$ s$^{-1}$ in the in-plane direction parallel to the base 'W', which is perpendicular to the height 'H' of the pre-existing crack during uniaxial tensile loading. The system was maintained at room temperature by using Nosé-Hoover thermostat and barostat [100,101]. Number of particles in the system was maintained constant. For biaxial tensile loading, tension was applied along both axes (armchair and zigzag directions) concurrently using the same strain rate $10^8$ s$^{-1}$. This strain rate $10^8$ s$^{-1}$ is commonly employed in atomistic simulations due to its computational efficiency while maintaining reasonable accuracy in capturing TMD material failure events [69,72,75]. In all simulations, we employed a 1 fs timestep to integrate atomic motion and compute atomic stresses using the virial stress theorem [102].

## 4. Stress-Strain Relationship:

The atomic stress-strain behavior was studied by applying uniaxial deformation to the simulation cell and then averaging the stress across the whole structure. For the biaxial tensile loading, the simulation box was subjected to deformation in both axes (X and Y axes) concurrently. To calculate atomic stress, the virial stress approach was employed. Virial stress components were calculated using the expression in Eq. 1 [103]:

$$\sigma_{Virial}(r) = \frac{1}{\Omega} \sum_i [(-m_i \dot{u}_i \otimes \dot{u}_i + \frac{1}{2} \sum_{j \neq i} r_{ij} \otimes f_{ij})] \qquad (1)$$

The summation was considered over all atoms present within the entirety of volume $\Omega$. $\otimes$ symbolizes the cross-product. The mass of atom 'i' is denoted by $m_i$. This atom's position vector is $r_{ij}$. The time derivative $\dot{u}_i$ indicates the change in the position of an atom in relation to a predetermined reference location. $f_{ij}$ represents the interatomic force that atom 'j' exerts on atom 'i'. As the stress is calculated without taking into account the actual thickness of the structure, the N/m unit is used.

The formula in Eq. 2 was used to compute strain:

$$\varepsilon = \frac{L_x - L_{x0}}{L_{x0}} \tag{2}$$

$L_{x0}$ and $L_x$ variables in Eq. 2 represent the original undeformed initial length and the instantaneous length of the simulation box, respectively.

The accuracy of MD simulations heavily relies on the choice of interatomic potential. In this study, we made use of the Stillinger-Weber (SW) potential, which has been proven effective in characterizing the mechanical properties of 2D monolayer transition metal dichalcogenides (TMDs) [71,104–107]. Specifically, we adopted the SW potential developed by Jiang [76], which has been widely utilized in various studies to predict the mechanical properties of many 2D materials [73,108]. The mathematical expressions for corresponding terms are as follows:

$$\phi = \sum_{i<j} V_2 + \sum_{i>j<k} V_3 \tag{3}$$

$$V_2 = A e^{\left[\frac{\rho}{r-r_{max}}\right]} \left(\frac{B}{r^4} - 1\right) \tag{4}$$

$$V_3 = K\varepsilon e^{\frac{\rho_1}{r_{ij}-r_{maxij}} \frac{\rho_2}{r_{ik}-r_{maxik}}} (\cos\theta - \cos\theta_0)^2 \tag{5}$$

$V_2$ and $V_3$ were used to refer to the bond-stretching and angle-bending terms respectively. $r_{max}$, $r_{maxij}$, and $r_{maxik}$ parameters refer to the cutoff values that play a crucial role in our calculations. $\theta_0$ symbol denotes the angle formed between two bonds in their equilibrium state. Energy-related parameters A and K were determined from the Valence Force Field (VFF) model. Additionally, we have considered other parameters such as $B, \rho, \rho_1$ and $\rho_2$. These parameters serve as fitted coefficients. The specific values for these parameters are obtained from reference [76].

## 5. Results and Discussions

### 5.1. Uniaxial Tension:

All the structures we are studying were subjected to uniaxial tensile loading to determine the stress-strain relationship, which is essential to characterize the mechanical properties of a material. Initially, tensile loading simulations were performed on monolayer MoTe$_2$ with triangular pre-crack defects with almost the same porosity, under $10^8$ s$^{-1}$ strain rate at room temperature (300K). To study the effect of vertex angles of the triangle-shaped vacancy defects in MoTe$_2$, the strain was applied on the axis parallel to the base 'W' of the isosceles-shaped triangular vacancy defect of that system. It means if the isosceles triangle base 'W' is positioned along the X-axis and height 'H' is positioned along the Y-axis (for example X60° system), uniaxial stress is applied along the X-axis and vice versa. In other words, X30°, X60°, X75°, X90°, X105°, X120°, X150° systems were subjected to strain along the X-axis (armchair orientation) (**Figure 3**(a)), while Y30°, Y60°, Y75°, Y90°, Y105°, Y120°, Y150° systems were subjected to strain along the Y-axis (zigzag orientation) (**Figure 3**(c)). The stress-strain curves resulting from uniaxial tensile loading along the X-axis (armchair direction uniaxial tensile loading) are presented in **Figure 3**(b), while those for the uniaxial tensile loading along the Y-axis (zigzag direction uniaxial tensile loading) are shown in **Figure 3**(d).

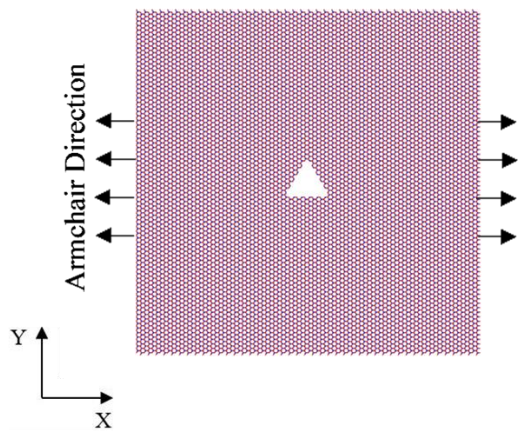
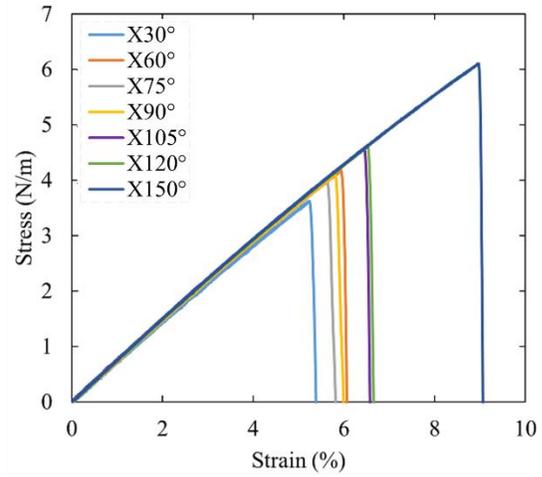
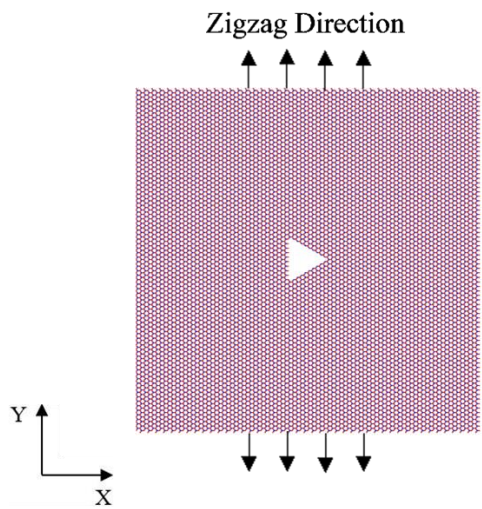
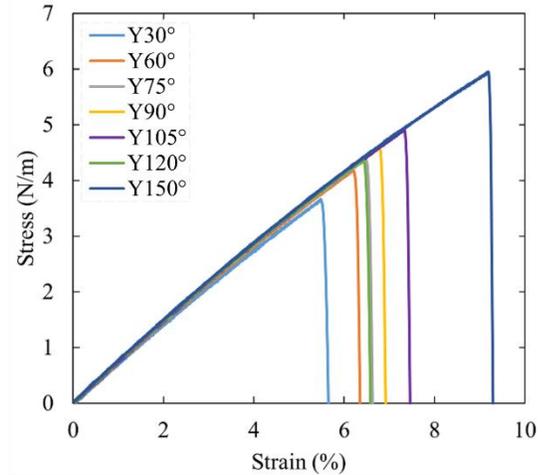

***Figure 3*** *(a) Uniaxial tensile loading of an "X_°" system, where tensile strain is applied in the armchair direction. (b) Resulting uniaxial stress-strain relationship of X30°, X60°, X75°, X90°, X105°, X120°, X150° systems. (c) Uniaxial tensile loading of a "Y_°" system, where tensile strain is applied in the zigzag direction. (d) Resulting uniaxial stress-strain relationship of Y30°, Y60°, Y75°, Y90°, Y105°, Y120°, Y150° systems.*

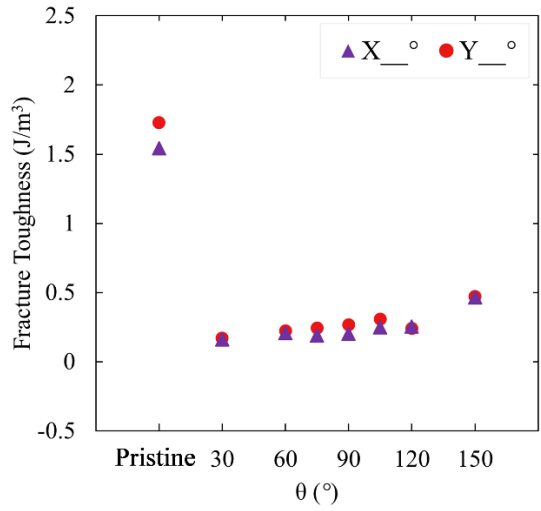 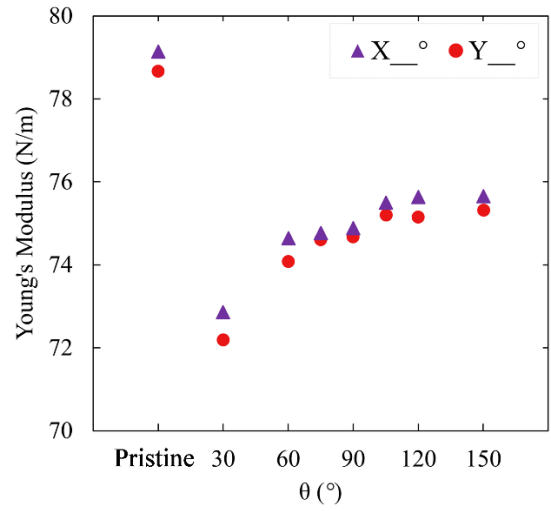
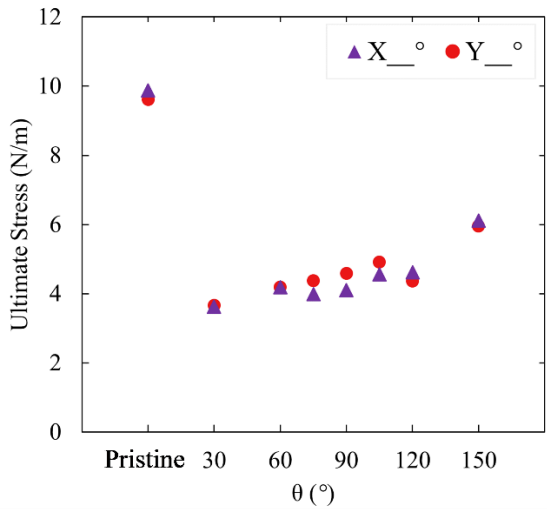 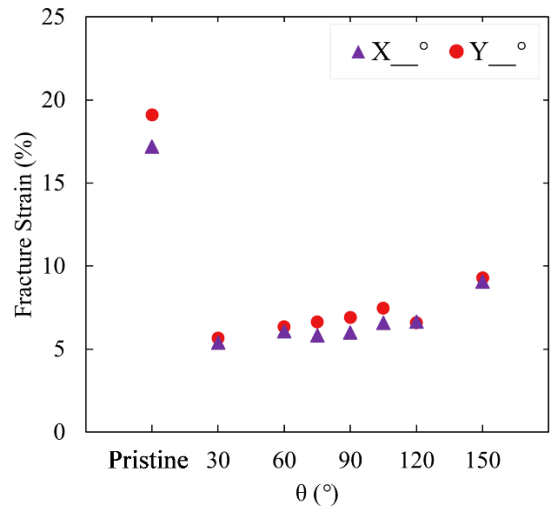

*Figure 4* Impact of vertex angle of the triangle-shaped pre-crack on the mechanical properties of pre-cracked monolayer MoTe$_2$: (a) fracture toughness, (b) Young's modulus, (c) ultimate tensile strength, and (d) fracture strain under uniaxial tensile loading.

The stress-strain curves observed in all fourteen systems exhibited a common trend: an initial linear rise in stress values with increasing strain, eventually reaching a point where it experiences abrupt fracture. This behavior suggests that all fourteen systems are brittle. Furthermore, it corroborates earlier findings that failure in typical pre-cracked 2D materials is brittle [70,109–111]. For the uniaxial tensile loading of "X_°" systems, failure is noticed to occur in the following order – X30°, X75°, X90°, X60°, X105°, X120°, X150°. When subjected to uniaxial tensile loading, "Y_°" systems fail in this order - Y30°, Y60°, Y120°, Y75°, Y90°, Y105°, Y150°. Pre-crack length [70,107,111–113], width [112], orientation [114,115], position [70] as well as tip configuration [116] significantly influence the fracture properties of these systems. As the porosity remained nearly constant in all systems, an increase in the vertex angle led to a decrease in pre-crack length (for details see supplementary information **Figure S 1** and **Figure S 2**). Longer pre-crack causes 2D materials to fail early. In this study, we have assessed key material properties such as Young's modulus, ultimate tensile strength, fracture strain, and toughness by analyzing the stress-strain relationships for each system. Fracture toughness is indicative of the useful strength of a material. Fracture toughness was determined by measuring the area under the stress-strain curve from the origin to the fracture point [117]. By linearly fitting the stress-strain relation in the short strain range of [0, 0.01], Young's modulus was determined [76]. Young's modulus is also known as elastic modulus, and quantifies the material's stiffness. Ultimate tensile strength or ultimate stress indicates the maximum amount of tensile stress a material can endure before it undergoes failure. The ultimate tensile strength can be determined by identifying the maxima on the corresponding stress-strain curve. These mechanical properties are illustrated in **Figure 4**. Systems with vertex angles 150°, X150° and Y150° had shown higher ultimate tensile strength and fracture strain than other "X_°" systems and "Y_°" systems during armchair direction uniaxial tensile loading and zigzag direction uniaxial tensile loading, respectively. Previous studies have found that for both armchair direction uniaxial tensile loading and zigzag direction uniaxial tensile loading, ultimate stress and failure strain deteriorate with the increase in pre-crack length for linear pre-crack placed in zigzag and armchair orientation, respectively [70,111]. In the systems we are studying, the pre-crack length decreases when the vertex angles of the triangle-shaped pre-crack increase. Systems with vertex angles 150°, X150°, and Y150° possess the lowest pre-crack lengths and largest vertex angles (see **Table 1** and supplementary information **Figure S 1**, **Figure S 2**) in their respective categories. Similarly, systems with vertex angles of 30°, X30°, and Y30° having the lowest vertex angles and highest pre-crack lengths (see **Table 1** and supplementary information **Figure S 1**, **Figure S 2**) had demonstrated the lowest ultimate tensile strength and fracture strain in their respective categories. Hence, the material properties of X30°, X150°, Y30°, and Y150° are in sync with the conclusion that increasing pre-crack length deteriorates mechanical properties. With nearly identical percentage of defect, X150° system exhibits 3.84% higher Young's modulus, 68.32% greater fracture strain, 68.43% higher ultimate tensile strength, and 190.37% higher fracture toughness compared to X30° system under armchair direction uniaxial tensile loading. Similarly, Y150° system demonstrates 4.32% higher Young's modulus, 64.44% greater fracture strain, 62.37% higher ultimate tensile strength, and 175.99% higher toughness compared to the Y30° system under zigzag direction uniaxial tensile loading. X30° system fails at 5.39% strain, where the corresponding ultimate stress is 3.63 N/m. This system has a toughness of 0.16 J/m$^3$. On the other hand, the X150° system fails at 9.06% strain, where the corresponding ultimate stress is 6.11 N/m applied along the X-axis (armchair orientation), demonstrating a toughness of 0.46 J/m$^3$. Among "Y_°" systems, where stress was applied along the Y-axis (zigzag orientation), Y30° failed at 5.65%, where corresponding ultimate stress is 3.67 N/m and toughness is 0.17 J/m$^3$. For this Y30° system, Young's modulus, fracture strain, ultimate stress and toughness values are lowest among the "Y_°" systems subjected to zigzag direction uniaxial tensile loading. Y150°, on the other hand, resists failure till 9.29% strain when applied stress reaches 5.95 N/m during zigzag direction uniaxial tensile loading, having a toughness of 0.47 J/m$^3$. For this Y150° system, all the parameters we are studying exhibit greater values than other "Y_°" systems in zigzag direction uniaxial tensile loading. Contrary to this pattern, it is noticed from the obtained results that for the armchair direction uniaxial tensile loading of "X_°" systems, X60° outperforms X75° and X90° in terms of ultimate stress, fracture strain, and toughness, even though the later systems have a pre-crack length smaller than the former one (see supplementary information **Figure S 1**). For zigzag direction uniaxial tensile loading of "Y_°" systems, Y75°, Y90° and Y105° outperform Y120° in terms of ultimate stress, fracture strain, and toughness.

For both "X_°" and "Y_°" systems, Young's modulus has increased when the vertex angle has increased, without any exception. Among the "X_°" systems, X150° possesses the highest Young's modulus of 75.66 N/m, which is about 4.39% less than the Young's modulus of pristine MoTe$_2$, when this pristine MoTe$_2$ is subjected to armchair direction uniaxial tensile loading. As the pre-crack angle decreases, so does the Young's modulus. X30° system has the lowest Young's modulus value of 72.86 N/m among "X_°" systems. Similarly, for the zigzag direction uniaxial tensile loading of "Y_°" systems, Y150° demonstrates the highest resistance to failure with Young's modulus of 75.32 N/m, followed by Y105°, Y120°, Y90°, Y75°, Y60° and Y30° systems. Among these "Y_°" systems, the Y30° system has the lowest Young's modulus of 72.19 N/m, which is 8.23% lower than Young's modulus of pristine MoTe$_2$, when this pristine MoTe$_2$ is subjected to zigzag direction uniaxial tensile loading.

Overall, for monolayer MoTe$_2$ with triangle-shaped pre-cracks, an association can be observed between the relationship of the height and base length of the triangle-shaped pre-cracks and the subsequent influence of this relationship on the mechanical properties resulting from uniaxial tensile loading (see supplementary information **Figure S 1**, **Figure S 2**). Monolayer MoTe$_2$ with triangle-shaped pre-cracks becomes less resistant to failure as the vertex angle decreases, leading to pre-cracks that increasingly resemble conventional straight cracks with a prominent height relative to the base length. On the other hand, an increase in the vertex angle enhances failure resistance when the pre-crack more closely resembles a straight crack with reduced height compared to the base length.

**5.2. Biaxial Tension**

Following the uniaxial tensile loading, both "X_°" and "Y_°" systems were subjected to the biaxial tensile loading. During the biaxial tensile loading, the strain was applied in both X and Y directions (armchair and zigzag orientations, respectively) concurrently (**Figure 5((a),(c))**). The impact of vertex angles in triangle-shaped vacancy defects on the mechanical properties of monolayer $MoTe_2$ was studied through biaxial tensile tests, with the systems having nearly identical porosity levels. Tensile tests were performed under a strain rate of $10^8$ $s^{-1}$ at room temperature (300K). Material failure is deemed to occur if it fails in either armchair axis or zigzag axis. From the stress-strain relationship (**Figure 5((b),(d))**) obtained, we calculated material properties such as Young's modulus, fracture strain, ultimate tensile strength, and toughness for comparative analysis. The comparison is shown in **Figure 6**. Both pristine $MoTe_2$ and $MoTe_2$ systems with triangle-shaped pre-cracks were found to demonstrate higher Young's modulus when subjected to biaxial tensile

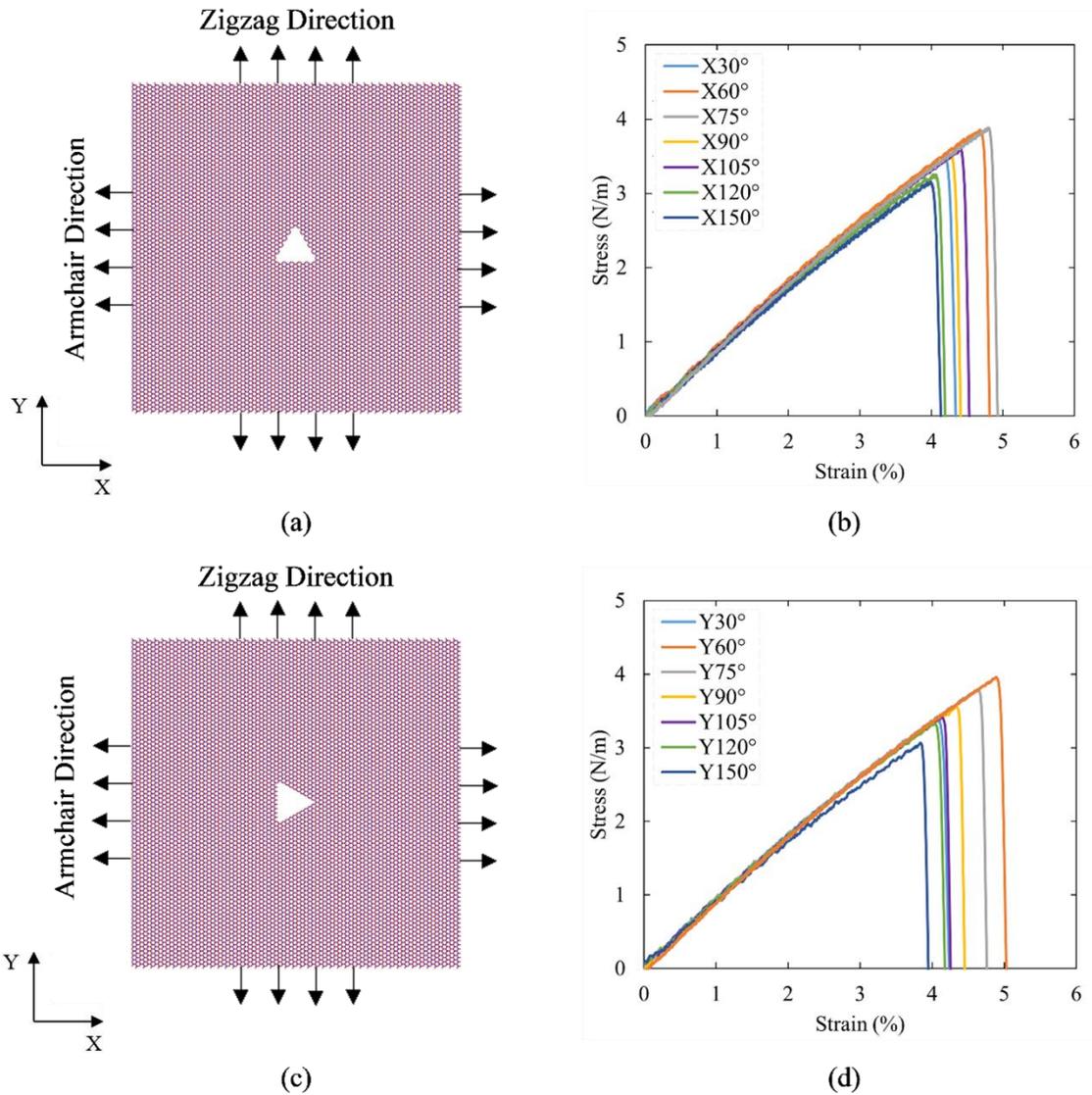

*Figure 5* *(a) Biaxial tensile loading of a "X_°" system, where tensile strain is applied concurrently in the armchair and zigzag directions. (b) Resulting biaxial stress-strain relationship of X30°, X60°, X75°, X90°, X105°, X120°, X150° systems. (c) Biaxial tensile loading of an "Y_°" system, where tensile strain is applied concurrently in the armchair and zigzag directions. (d) Resulting uniaxial stress-strain relationship of Y30°, Y60°, Y75°, Y90°, Y105°, Y120°, Y150° systems.*

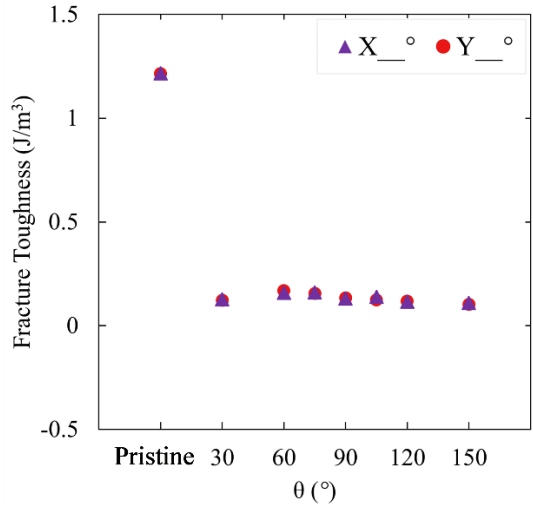 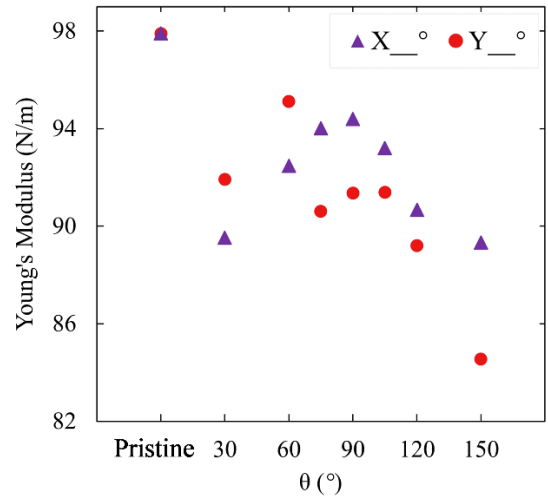

(a) (b)

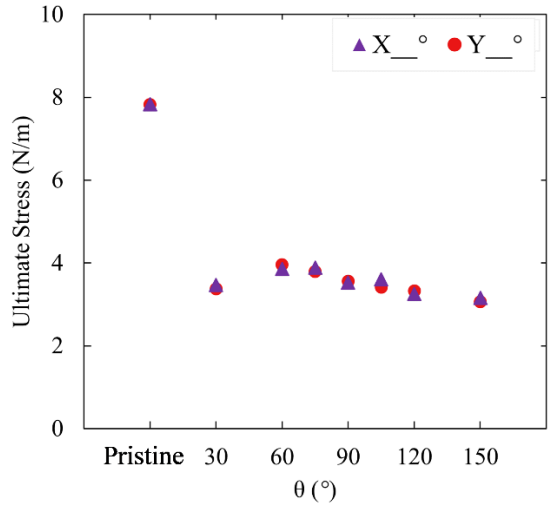 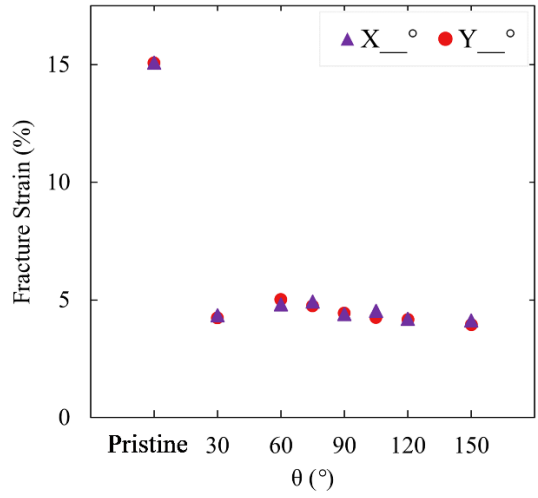

(c) (d)

*Figure 6* *Impact of vertex angle of the triangle-shaped pre-crack on the mechanical properties of pre-cracked monolayer MoTe$_2$: (a) fracture toughness, (b) Young's modulus, (c) ultimate tensile strength, and (d) fracture strain under biaxial tensile loading.*

loading (see **Figure 6**) compared to when they were subjected to the uniaxial tensile loading (see **Figure 4**). The Young's modulus obtained from biaxial tensile loading for X30°, X60°, X75°, and

X90° systems exhibited an ascending trend, measuring 89.52 N/m, 92.47 N/m, 94.01 N/m, and 94.39 N/m, respectively. X90° has the highest Young's modulus among the "X_°" systems. From there, Young's modulus decreases to 93.2 N/m for the X105° system and continues decreasing for the X120° and X150° systems. X120° and X150° systems demonstrate Young's modulus of 90.66 N/m and 89.33 N/m, respectively, under biaxial tensile loading. Young's modulus resulting from biaxial tensile loading of "Y_°" systems, too, does not show any particular trend. In the biaxial tensile loading, loading is applied along both armchair and zigzag directions concurrently. Hence, the vertex angle, one of the three angles that a triangle-shaped pre-crack has, alone does not influence the material failure properties. Initially, Young's modulus under biaxial tensile loading rises from the Y30° system to the Y60° system, from 91.92 N/m to 95.11 N/m. Then, for Y75° system, Young's modulus declines to 90.62 N/m. Subsequently, it rises to 91.36 N/m in the Y90° system, increases further to 91.39 N/m in the Y105° system, decreases to 89.19 N/m in the Y120° system, and finally drops to 84.55 N/m in the Y150° system. Young's modulus under biaxial tensile loading of pristine structure is higher than both "X_°" and "Y_°" systems with pre-existing cracks. The Young's modulus of pristine $MoTe_2$ is 97.89 N/m, resulting from biaxial tensile loading. In the uniaxial tensile loading of both "X_°" and "Y_°" systems, it was observed that as the vertex angle increased in value, Young's modulus also increased. Young's modulus resulting from the biaxial tensile loading of the identical systems did not show that trend observed under uniaxial tensile loading.

However, when fracture strain, ultimate tensile strength, and fracture toughness calculated from the stress-strain relationship of biaxial tensile loading were plotted against the perimeter of the triangle-shaped pre-cracks (**Figure 7**), patterns emerged for both "X_°" and "Y_°" systems.

Among "Y_°" systems, Y150° has the largest pre-crack perimeter, approximately 12.496 nm.

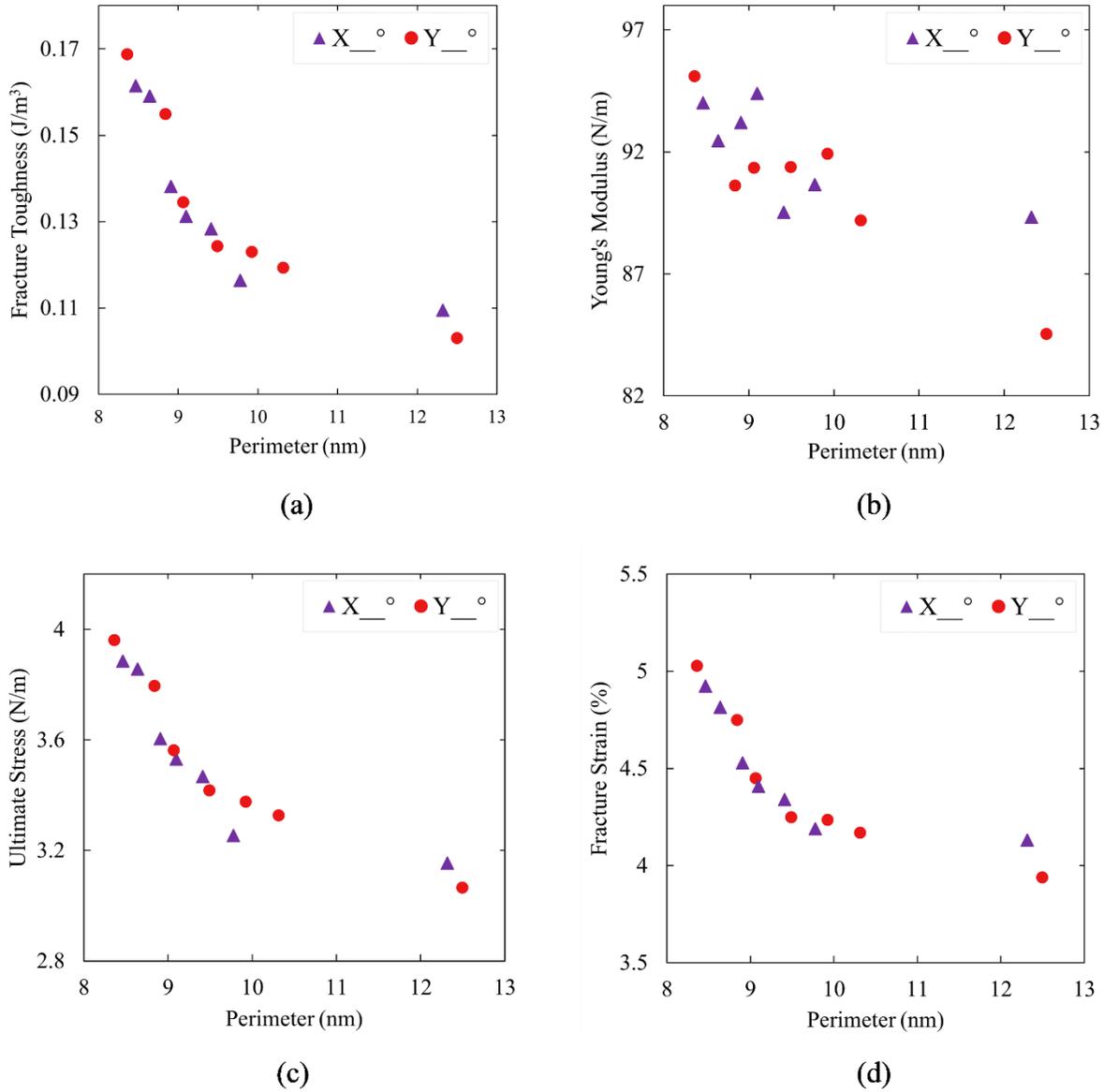

*Figure 7 Impact of the pre-crack perimeter on the mechanical properties of pre-cracked monolayer MoTe$_2$: (a) fracture toughness, (b) Young's modulus, (c) ultimate tensile strength, and (d) fracture strain under biaxial tensile loading.*

Under biaxial tensile loading, the Y150° system's fracture strain is 3.94%, and ultimate stress is 3.07 N/m. All these properties exhibit lower values compared to other "Y_°" systems with smaller perimeters. Y60°, having the smallest perimeter among "Y_°" systems, outperforms other "Y_°" systems when subjected to biaxial tensile loading. Y60° system fails when strain reaches 5.03%. Its ultimate stress and fracture toughness values are 3.96 N/m and 0.17 J/m$^3$ respectively, both of which are highest among "Y_°" systems. With almost identical percentage of defect, Y150° system shows 27.61% higher fracture strain, 29.22% higher ultimate tensile strength, and 63.79% higher toughness when compared to the Y60° system under biaxial tensile loading. Similarly, for the biaxial tensile loading of "X_°" systems, X150°, having the largest pre-crack perimeter, fails at a lower strain compared to other "X_°" systems, at strain 4.13% with applied stress reaching 3.15 N/m. The X150° system's fracture toughness value is 0.11 J/m$^3$, lower than the toughness of other "X_°" systems with smaller pre-crack perimeter. When the pre-crack perimeter decreases, material properties resulting from the biaxial tensile loading, namely fracture strain, ultimate tensile strength, and toughness, improve. Among "X_°" systems, X75° has the lowest pre-crack perimeter. It outperforms other "X_°" systems with larger pre-crack perimeters. X75° reaches its ultimate stress of 3.89 N/m at 4.93% strain under biaxial tensile loading. It has a fracture toughness of 0.16 J/m$^3$ under biaxial tensile loading. With nearly identical defect percentages, X150° system exhibits 19.19% greater fracture strain, 23.18% higher ultimate tensile strength, and 47.53% higher toughness than X75° system under biaxial tensile loading. Since pristine MoTe$_2$ has no pre-crack, it outperforms all defective "X_°" and "Y_°" systems with pre-cracks under biaxial tensile loading. The fracture strain of pristine MoTe$_2$ is 15.08% under biaxial tensile loading. Its ultimate stress is 7.83 N/m, while its toughness is 1.22 J/m$^3$, as calculated from biaxial tensile loading stress-strain relationship. Fracture strain, ultimate stress, and fracture toughness parameters resulting from biaxial tensile loading are smaller than the values of the same parameters obtained from the uniaxial tensile loading of pristine MoTe$_2$.

It is evident that, for the biaxial tensile loading, fracture strain, ultimate tensile strength, and fracture toughness deteriorate when the pre-crack perimeter increases. In the same 'X_°' and 'Y_°' systems, uniaxial tensile loading yields superior fracture strain, ultimate tensile strength, and fracture toughness, but inferior Young's modulus compared to biaxial tensile loading. The Vector Displacement chapter explains why biaxial tensile loading of pre-cracked structures yields lower fracture strain, ultimate tensile strength, and toughness but higher Young's modulus compared to uniaxial tensile loading.

### 5.3. Fracture mechanism

For a comprehensive understanding of fracture mechanism, we have labeled the corners of the pre-cracks in both "X_°" and "Y_°" systems as "top corner", "left corner", and "right corner", illustrated in **Figure 8**. Subsequently, **Figure 9** and **Figure 10** depict the fracture mechanisms of "X_°" and "Y_°" systems under uniaxial tensile loading, while **Figure 11** and **Figure 12** illustrate the fracture mechanisms of these systems under biaxial tensile loading.

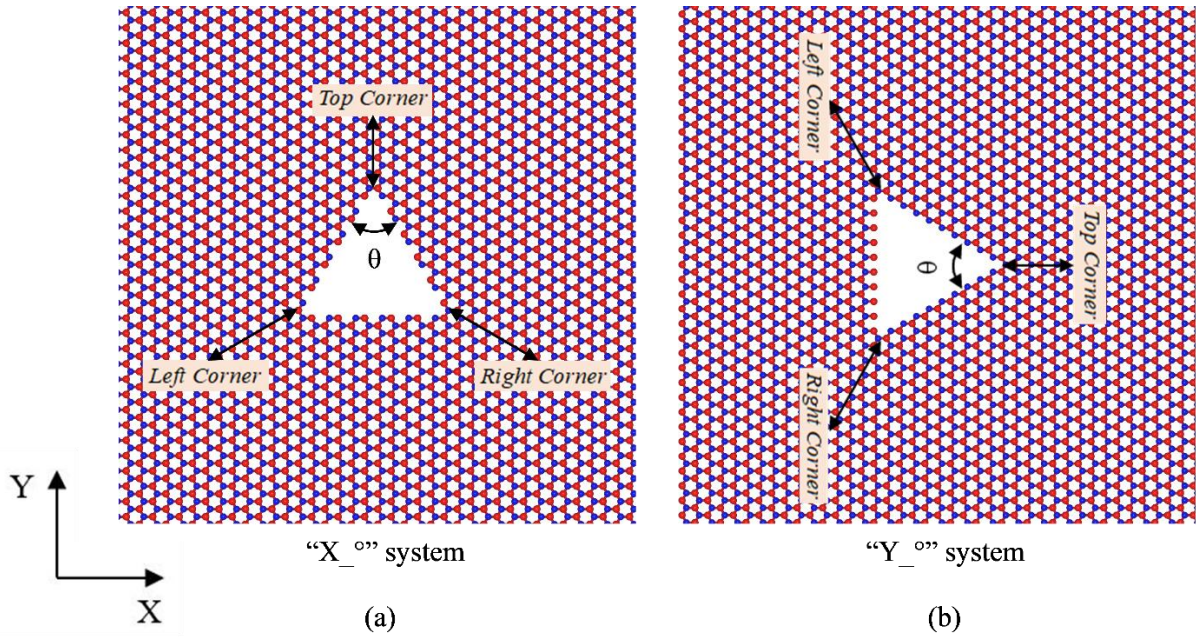

***Figure 8*** *In "X_°" and "Y_°" systems, three corners of a triangle-shaped pre-crack are referred as "Top corner", "Left corner" and "Right corner". The arrows indicate the name of the corners for an (a) "X_°" system and (b) "Y°" system. Both systems shown have vertex angles θ =60°.*

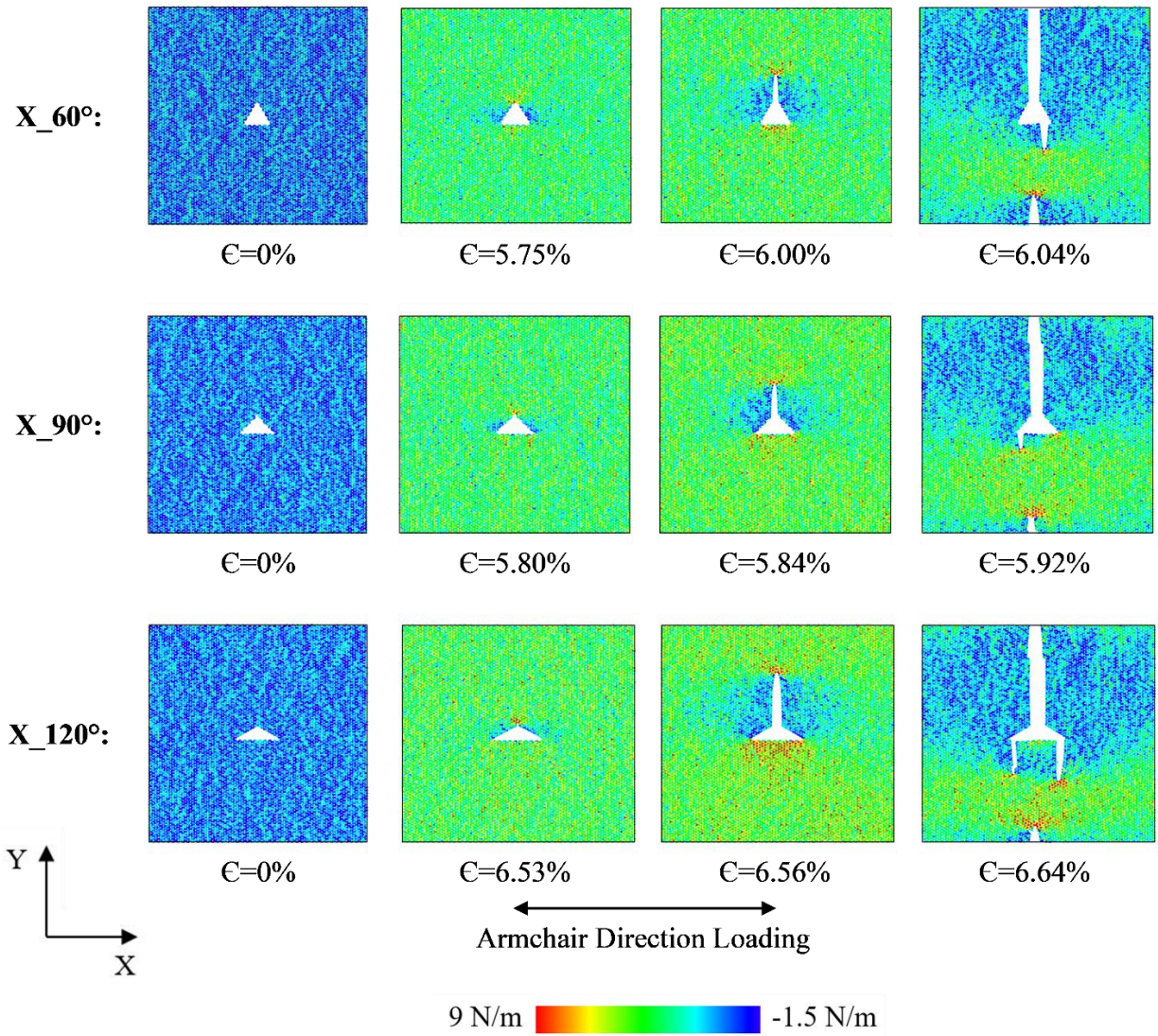

*Figure 9* Crack propagation and stress distribution during armchair direction uniaxial tensile loading of X_60°, X_90° and X_120° systems.

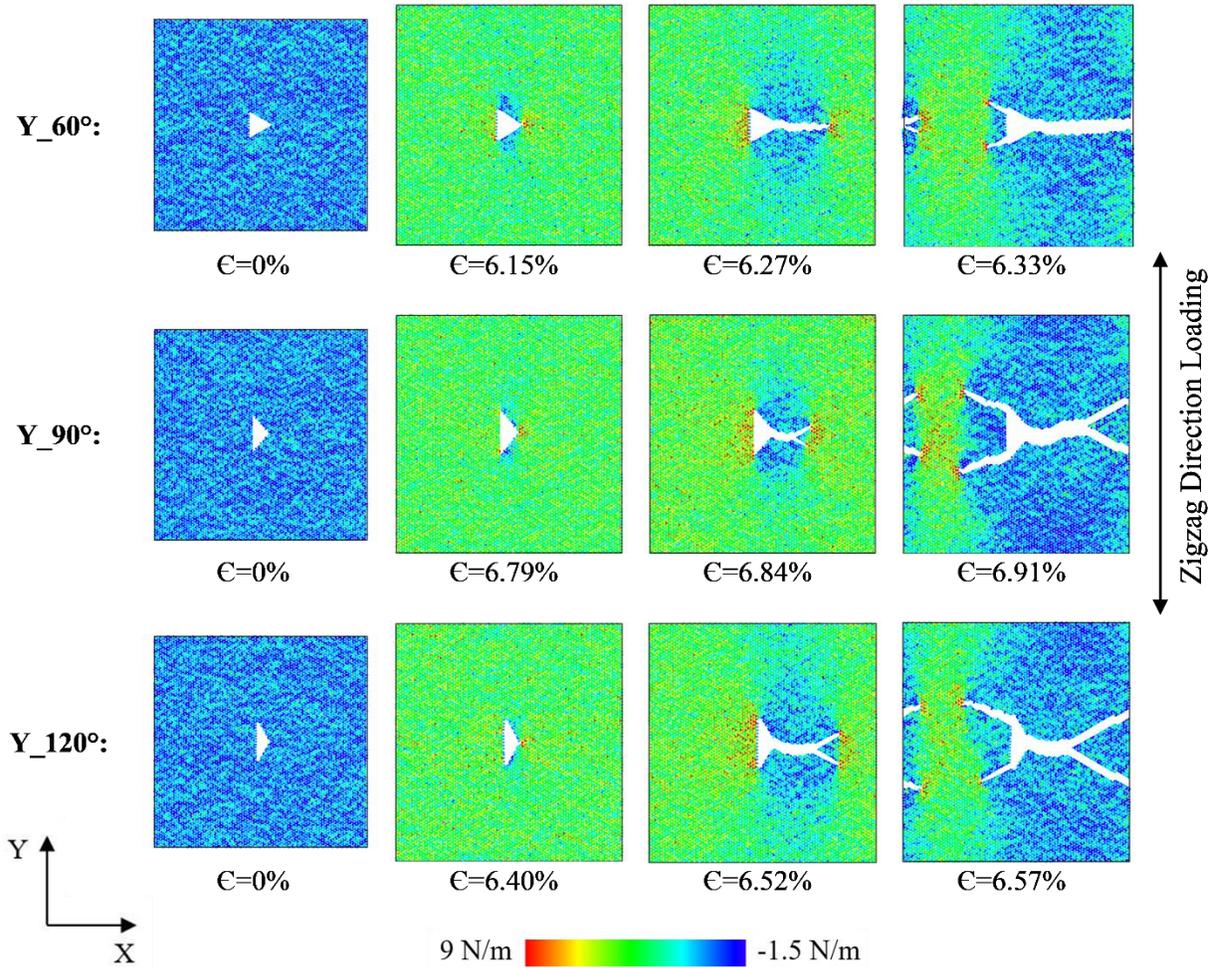

*Figure 10* Crack propagation and stress distribution during zigzag direction uniaxial tensile loading of Y_60°, Y_90° and Y_120° systems.

All the fracture visuals were created using the OVITO visualization tool [97]. During both uniaxial and biaxial tensile loading, with the increase of applied stress, we have observed local stress concentrating at the pre-crack corners. Crack propagates differently during armchair direction (X-axis) uniaxial tensile loading (see **Figure 9**) and zigzag direction (Y-axis) uniaxial tensile loading (see **Figure 10**).

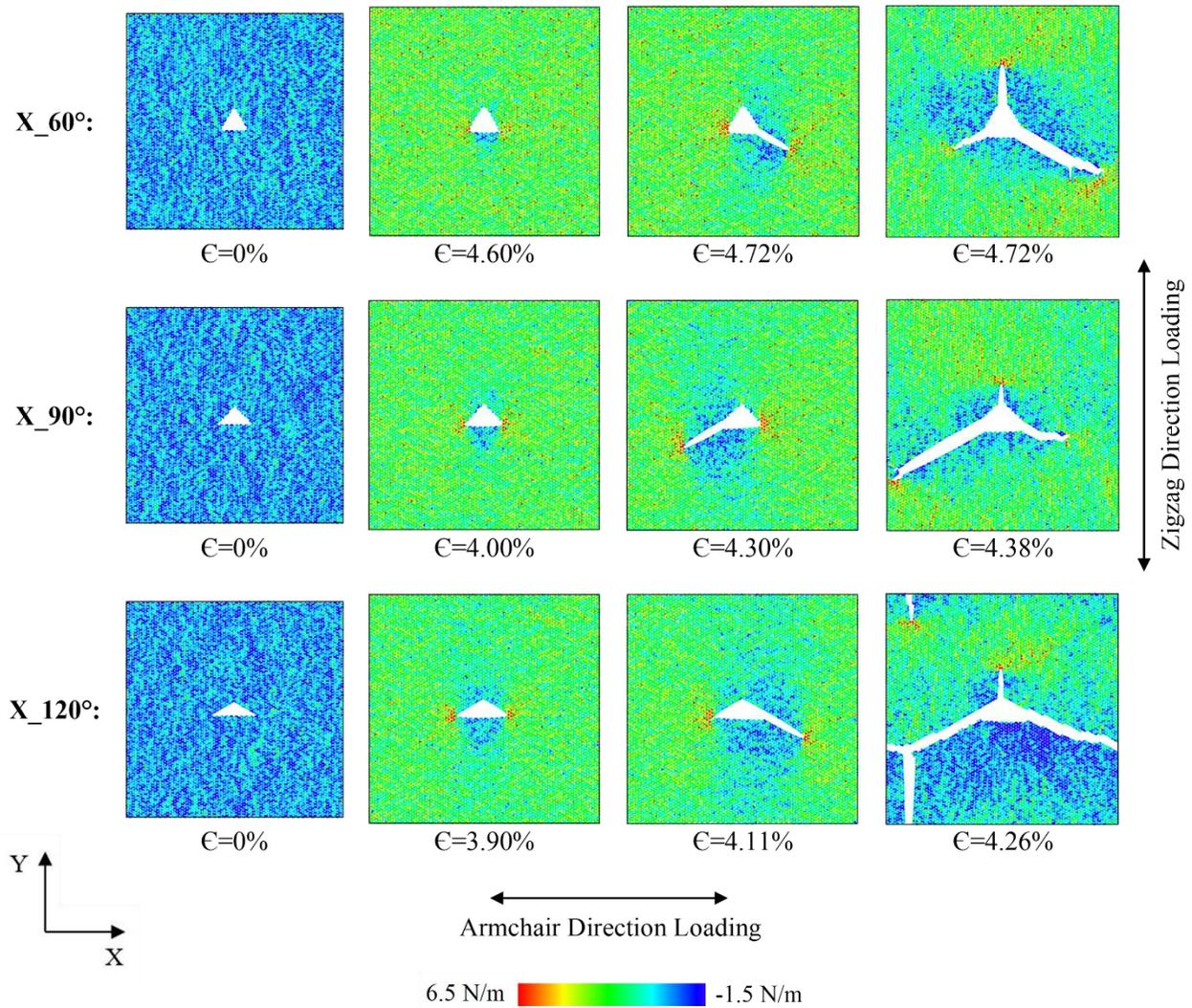

*Figure 11* Crack propagation and stress distribution during biaxial tensile loading of X_60°, X_90° and X_120° systems.

During zigzag direction uniaxial tensile loading of "Y_°" systems, two potential crack propagation routes emerge due to the symmetrically inclined bonds diverging at angles of ±60° with respect to the armchair direction (X-axis) (illustrated in supplementary information **Figure S 3**(c),(d)). Consequently, we observe branching phenomenon during zigzag direction uniaxial tensile loading (see **Figure 10**) [70,111,112].

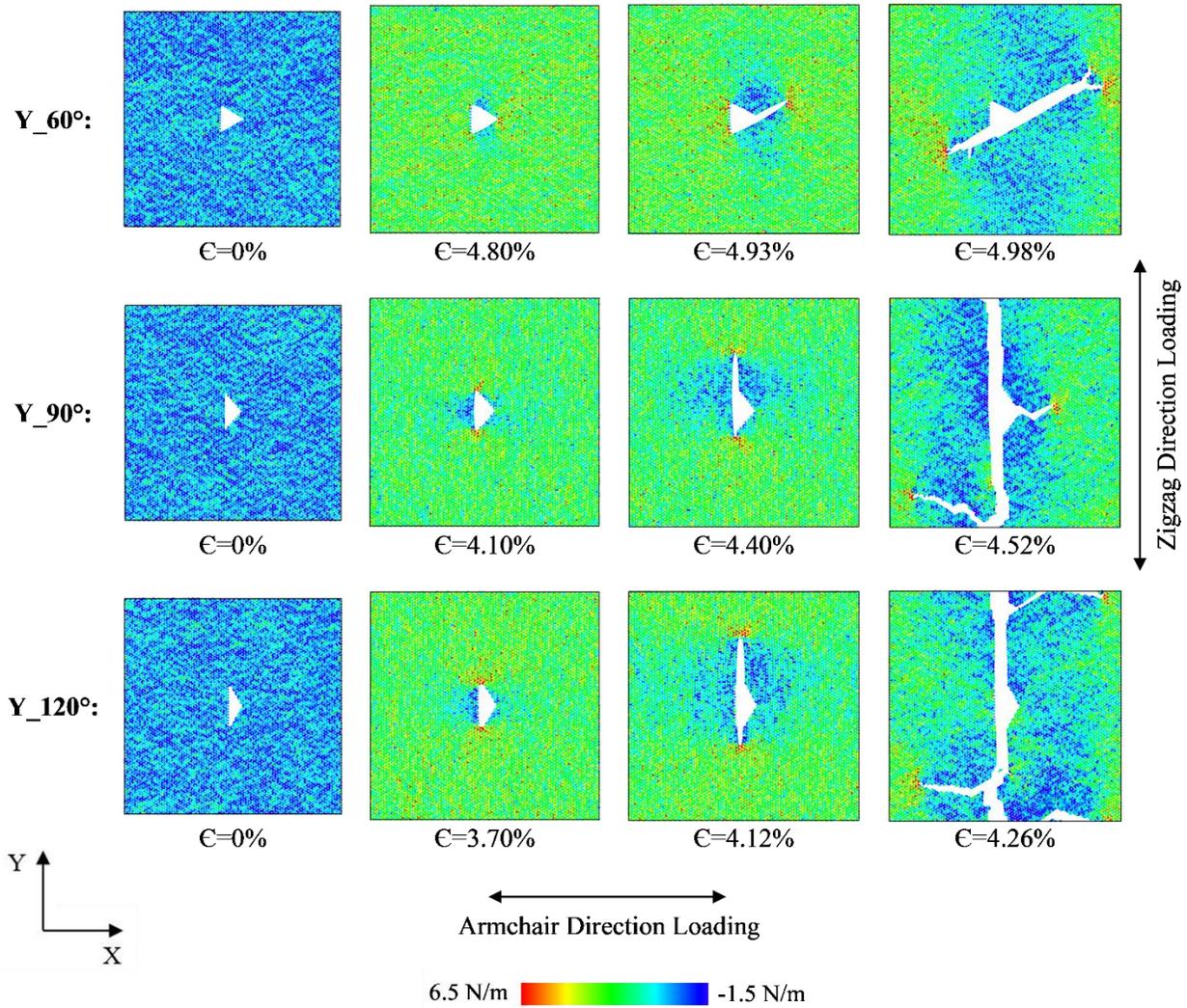

*Figure 12* Crack propagation and stress distribution during biaxial tensile loading of Y_60°, Y_90° and Y_120° systems.

On the other hand, for crack propagation along the Y-axis (zigzag orientation), bonds are positioned perpendicular to the direction of the pre-crack length 'H' from the point of view of the pre-crack tip (illustrated in supplementary information **Figure S 3**(a),(b)), providing a linear path for crack propagation. Hence, the crack nucleates from the vertex tip and propagates linearly along the zigzag direction (Y-axis) during armchair direction uniaxial tensile loading of "X_°" systems

(see **Figure 9**). This phenomenon highlights the differences in fracture mechanics along armchair and zigzag directions.

During the biaxial tensile loading, a different pattern in crack propagation is observed. For the convenience of explanation, we are going to refer to the pre-crack corners of both systems as the "top corner", "left corner", and "right corner" in a counterclockwise direction starting from the vertex tip (see **Figure 8**). In "X_°" systems crack starts to propagate in the right corner (**Figure 11** - X_60°, X120°) or left corner (**Figure 11** - X90°) and afterwards crack appears in remaining two corners including the top corner. For "Y_°" systems under biaxial tensile loading, crack first starts to propagate from any of the three corners (see **Figure 12**). If the crack first nucleates from the top corner (**Figure 12** - Y60°), then the crack propagates at an angle of 60° with respect to X-axis until branching. However, if the crack first appears in the left or right corner (**Figure 12**- Y90°, Y120°), this crack propagates along the Y-axis before crack appearing in the X-axis.

### 5.4. Vector Displacement

Atomic displacement vectors exhibit distinct patterns for uniaxial tensile loading (**Figure 13**(a),(c)) compared to biaxial tensile loading (**Figure 13**(b),(d)) in an identical system containing triangle-shaped pre-cracks. In uniaxial tensile loading of both "X_°" and "Y_°" systems, displacement vectors point inward from the base of the triangle-shaped pre-crack (**Figure 13**(a),(c)). In contrast, during biaxial tensile loading, displacement vectors extend outward from the base of the triangle-shaped pre-crack (**Figure 13**(b),(d)), further stretching the pre-crack outward. This outward stretching at the triangle's base in triangle-shaped pre-crack does not occur during uniaxial tensile loading.

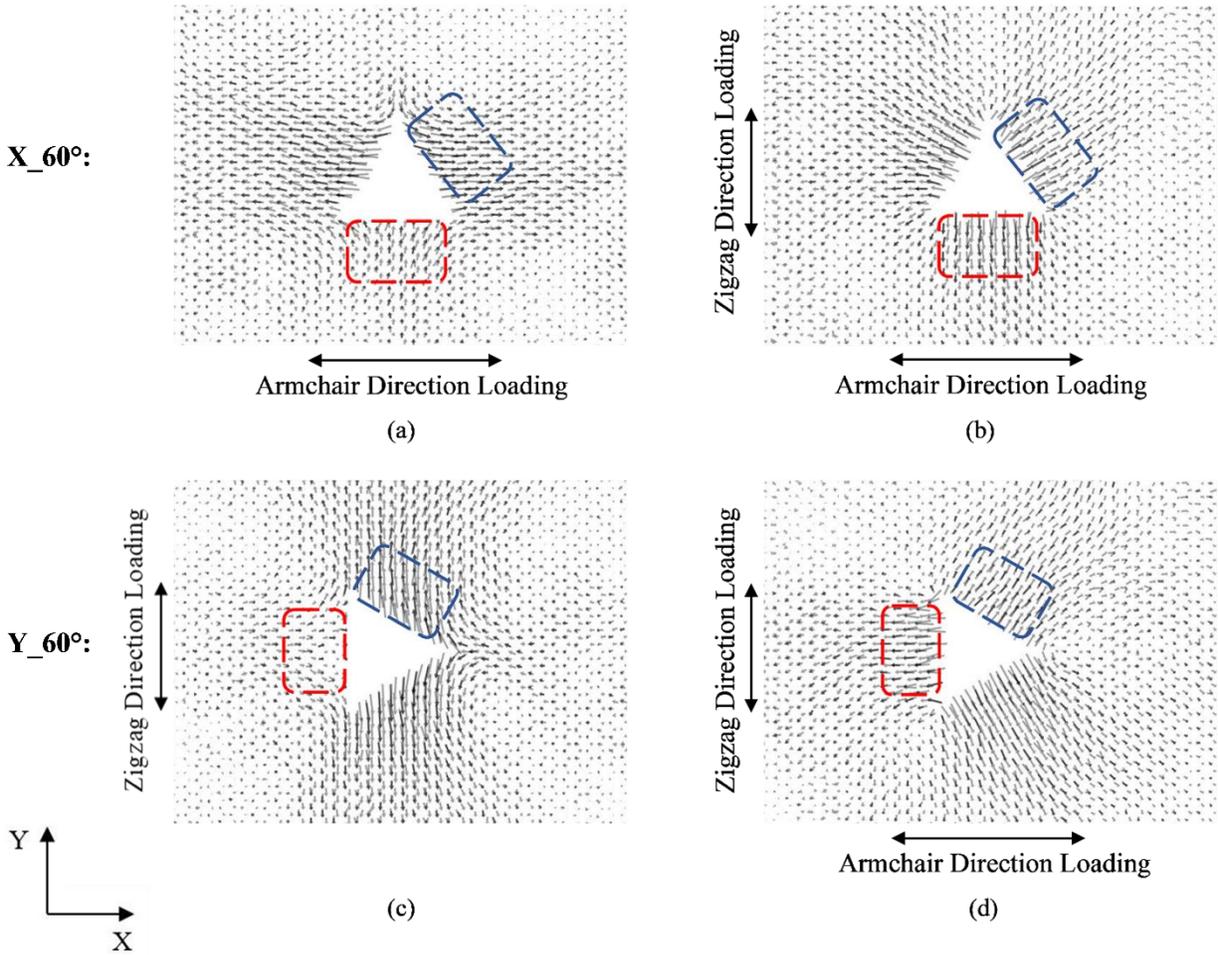

*Figure 13* Atomic displacement vectors around pre-cracks in (a,b) "X_°" and (c,d) "Y_°" systems under different loading conditions: (a) armchair direction uniaxial tensile loading, (b) biaxial tensile loading, (c) zigzag direction uniaxial tensile loading, and (d) biaxial tensile loading. Both 'X_°' and 'Y_°' systems shown in the figure have 60° vertex angles. The red and blue boxes enclose displacement vectors near the base and sides of the triangle-shaped pre-crack, respectively.

This results in lower fracture strain, ultimate tensile strength, and toughness for biaxial tensile loading than those values obtained under uniaxial tensile loading for identical systems.

During uniaxial tensile loading, displacement vectors align with the armchair orientation for a 'X_°' systems (**Figure 13**(a)) and with the zigzag orientation for 'Y_°' systems (**Figure 13**(c)), originating from the sides of the triangle-shaped pre-cracks. However, during biaxial tensile loading, the displacement vectors adjacent to the sides of the pre-crack triangle exhibit directions perpendicular to these sides (**Figure 13**(b),(d)), resulting in higher elasticity and improved Young's modulus.

## 6. Conclusion

In summary, our investigation focused on characterizing the mechanical and fracture properties of monolayer $MoTe_2$ containing triangle-shaped pre-existing cracks. We employed molecular dynamics simulations to evaluate these crucial mechanical and fracture attributes. The key points of our findings are listed below:

1. Mechanical properties could be manipulated by taking control of variable parameters, such as the pre-crack angle, shape, length and orientation.

2. Prior research has shown that an increase in pre-crack length generally results in reduction of fracture strain and ultimate tensile strength in TMDs [70]. This phenomenon is attributed to increased stress localization around the pre-crack tip with increasing pre-crack length [118], reducing ultimate stress and failure strain. However, our findings from uniaxial tensile loading simulations indicate that for a triangle-shaped pre-crack, pre-crack length alone may not impact the fracture behavior of TMDs. Nevertheless, we observed a consistent decrease in Young's modulus as the pre-crack length increased, accompanied by a reduction in the pre-crack vertex angle of monolayer $MoTe_2$ containing triangle-shaped pre-crack.

3. In biaxial tensile loading of monolayer $MoTe_2$ containing triangular pre-cracks, an increase in the pre-crack perimeter, while maintaining nearly constant porosity found, resulted in reduced fracture strain, ultimate tensile strength, and toughness.

4. Single layer MoTe$_2$ containing triangle-shaped pre-crack yields a higher Young's modulus under biaxial tensile loading than uniaxial tensile loading.

5. Single layer MoTe$_2$ demonstrates higher ultimate tensile strength, fracture strain, and toughness under uniaxial tensile loading, whether with or without defects, compared to biaxial tensile loading.

## 7. Data Availability

Data will be available on request.

## 8. Acknowledgment

## 9. Author Contributions

**M. J. Aziz:** Conceptualization, Simulation, Formal Analysis, Investigation, Data Curation, Visualization, Writing - Original Draft, Writing - Review & Editing, **M. A. Islam**: Conceptualization, Formal Analysis, Supervision, Writing - Review & Editing, **M.R. Karim**: Supervision, Writing - Review & Editing, **A. A. Bhuiyan**: Supervision, Writing - Review & Editing.

*Edition*, **2017**.

Supplementary Information of

# Effect of Triangular Pre-Cracks on the Mechanical Behavior of 2D $MoTe_2$: A Molecular Dynamics Study

Md. Jobayer Aziz[1], Md Akibul Islam[2*], Md. Rezwanul Karim[1], Arafat Ahmed Bhuiyan[1]

[1]Department of Mechanical and Production Engineering, Islamic University of Technology (IUT), Board Bazar, Gazipur 1704, Bangladesh

[2]Department of Mechanical and Industrial Engineering, University of Toronto, Canada

*Correspondence: aislam@mie.utoronto.ca (M.A.I.)

**Supplementary Figures**

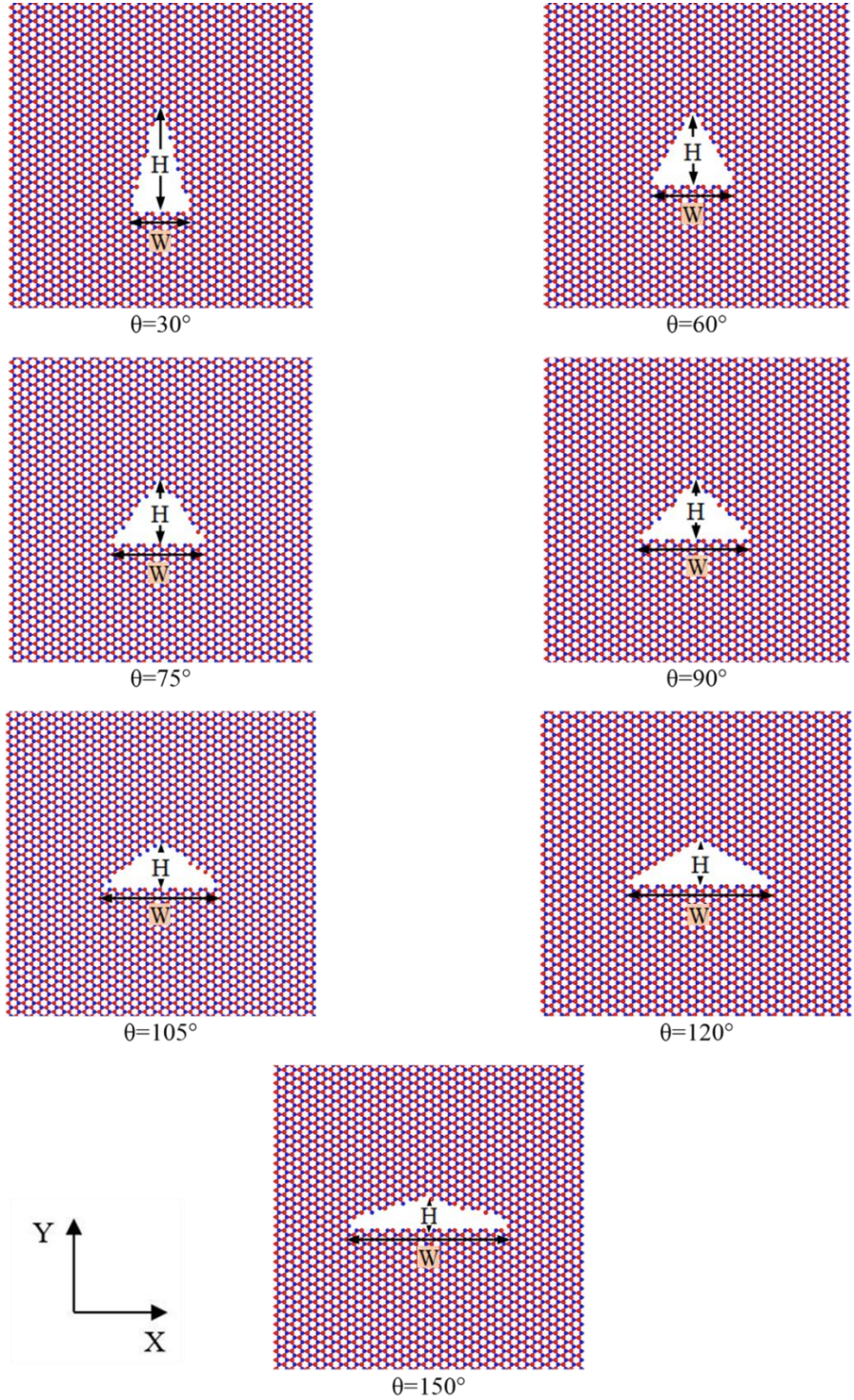

*Figure S 1* Atomic structure of MoTe$_2$ with triangle-shaped pre-cracks, where the base of the pre-crack 'W' is along the armchair axis (X-axis), while the height of the pre-crack triangle 'H' is

*along the zigzag axis (Y-axis). Such systems are referred as "X__°" systems with "__°" being the vertex angle (denoted by parameter 'θ') of the pre-crack triangle. With porosity nearly constant, increase in vertex angle 'θ' leads to reduction in the pre-crack length 'H'.*

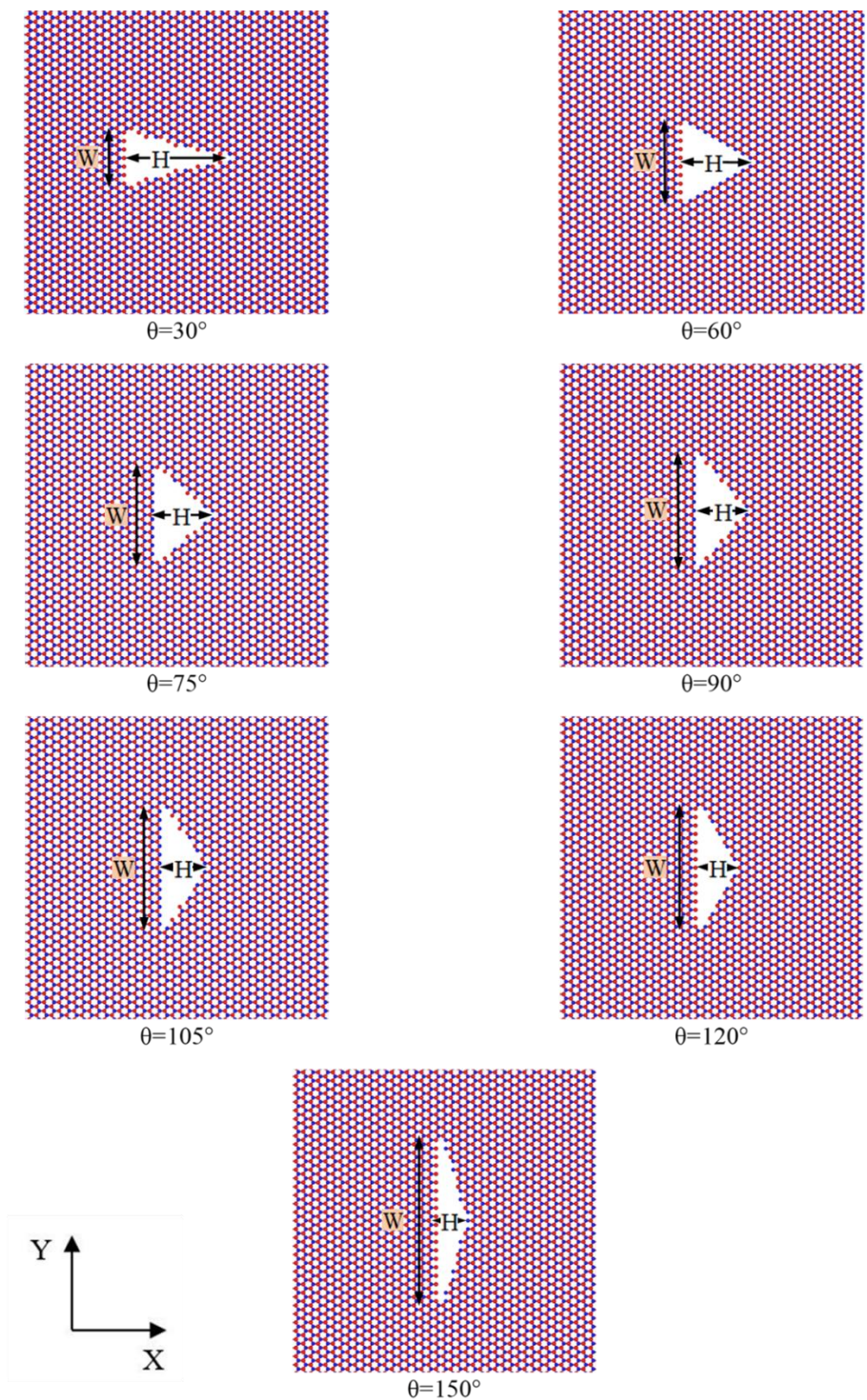

***Figure S 2*** *Atomic structure of MoTe$_2$ with triangle-shaped pre-cracks, where the base of the pre-crack 'W' is along the zigzag axis (Y-axis), while the height of the pre-crack triangle 'H' is along*

the armchair axis (X-axis). Such systems are referred as "Y__°" systems with "__°" being the vertex angle (denoted by parameter 'θ') of the pre-crack triangle. With porosity nearly constant, increase in vertex angle 'θ' leads to reduction in the pre-crack length 'H'.

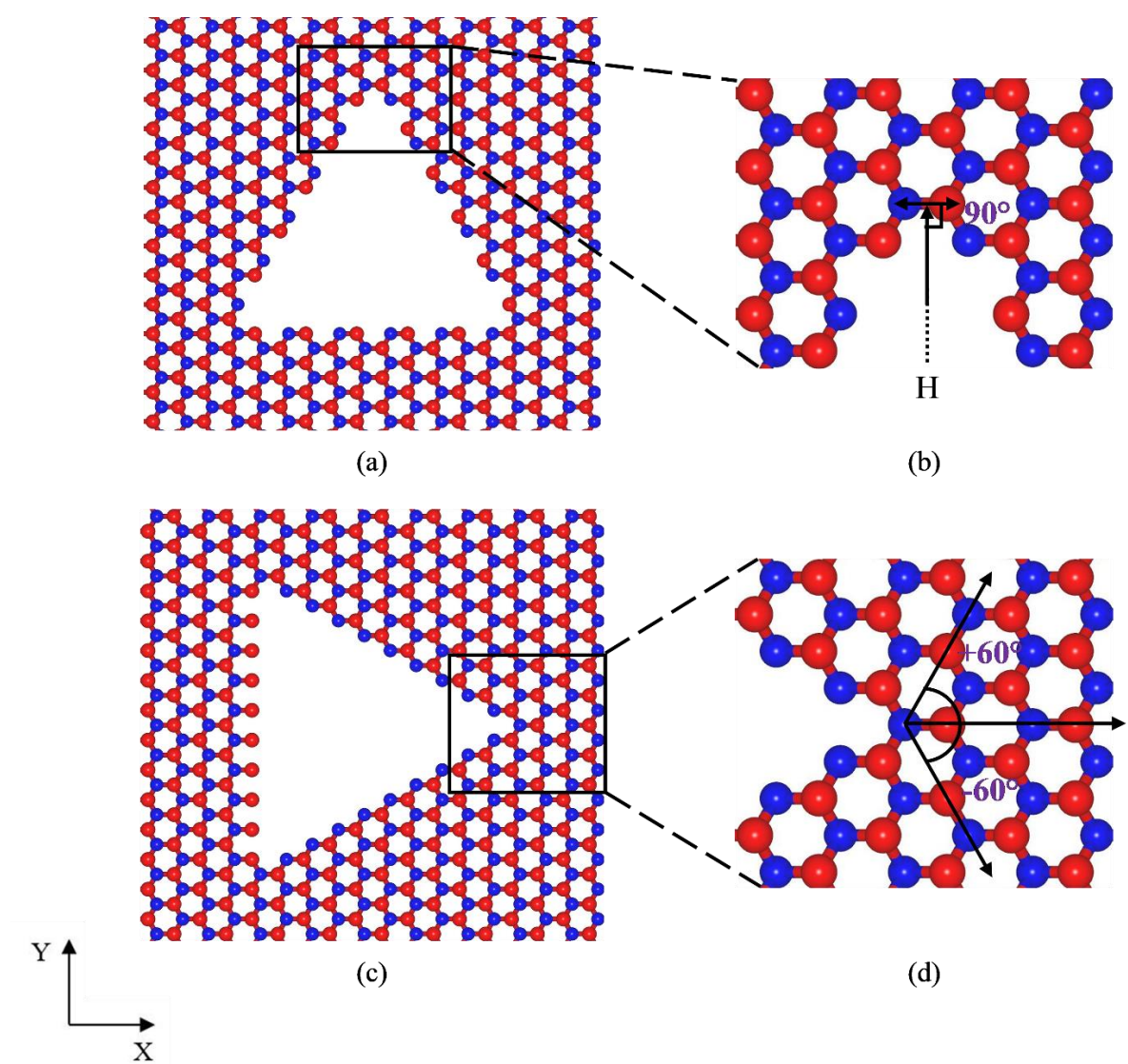

*Figure S 3* (a) Illustration of the triangle-shaped pre-crack in an "X_°" system where the base of the pre-crack triangle runs along the armchair axis (X-axis), and the pre-crack length aligns with the zigzag axis (Y-axis). (b) A zoomed-in view of the pre-crack tip in the "X_°" system reveals that

*the Mo-Te bonds are oriented perpendicular to the pre-crack's length 'H' in X-Y plane. (c) Illustration of the triangle-shaped pre-crack in a "Y_°" system where the base of the pre-crack runs along the zigzag axis (Y-axis), while the pre-crack length aligns with the armchair axis (X-axis). (d) A zoomed-in view of the pre-crack tip in the "Y_°" system shows bonds symmetrically inclined at ±60° angles relative to the armchair direction (X-axis) diverging from the pre-crack tip.*

**Method Validation**

In order to validate our approach, we subjected monolayer pristine 2H-MoTe$_2$ to uniaxial tensile loading in both the armchair and zigzag directions. We selected a 27 nm × 27 nm sheet for this study. The uniaxial tensile loading simulations were conducted at temperatures of 1K and 300K. The resulting stress-strain relationship is illustrated in ***Figure S 4***. From this dataset, we calculated essential material properties, including Young's modulus, ultimate tensile strength, and fracture strain. Comparing these findings with relevant literature values [1,2] in **Table S 1** reveals satisfactory agreement between our findings and existing research.

*Table S 1* *Validation of Young's modulus, ultimate tensile strength and fracture strain with existing literature.*

| | Current Study | | Ref [1] | | Ref [2] |
|---|---|---|---|---|---|
| | **Armchair loading** | **Zigzag loading** | **Armchair loading** | **Zigzag loading** | |
| **Young's Modulus (N/m)** | ~79.1397 | ~78.6709 | 79.8 | 78.5 | 79.4 |
| **Ultimate Tensile Stress (N/m)** | 11.664 (1K) 9.8823 (300K) | 11.136 (1K) 9.61086 (300K) | ~11.653 (1K) 9.945 (300K) | ~11.145 (1K) 9.819 (300K) | |
| **Fracture Strain (%)** | 0.256 (1K) 0.172 (300K) | 0.296 (1K) 0.191 (300K) | ~0.255 (1K) 0.173 (300K) | ~0.295 (1K) 0.198 (300K) | |

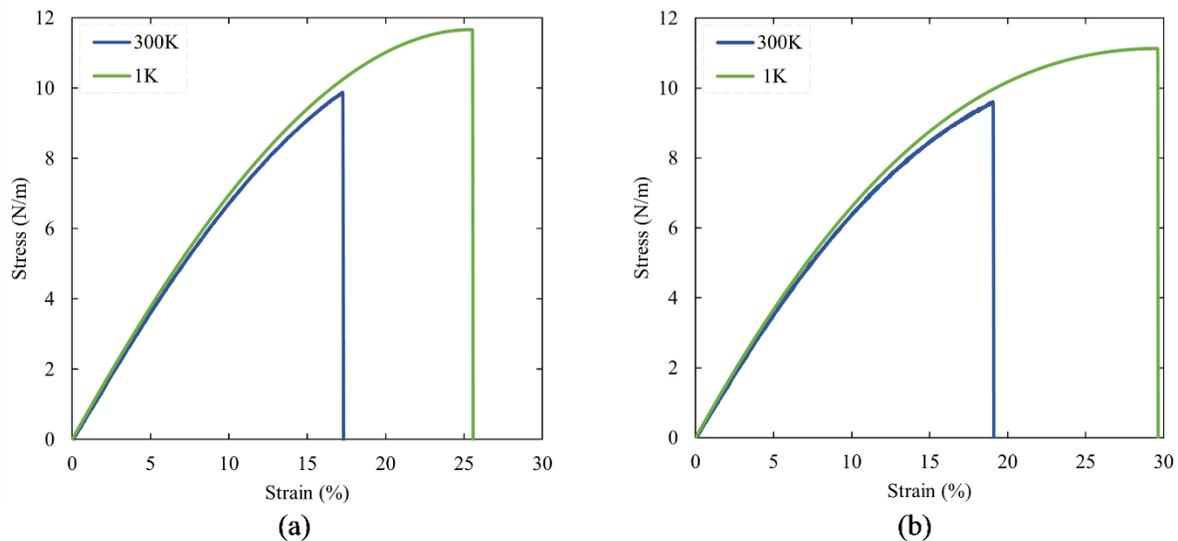

***Figure S 4*** *Stress-strain relationship of monolayer pristine 2H-MoTe$_2$ under (a) armchair direction uniaxial loading, (b) zigzag direction uniaxial loading.*

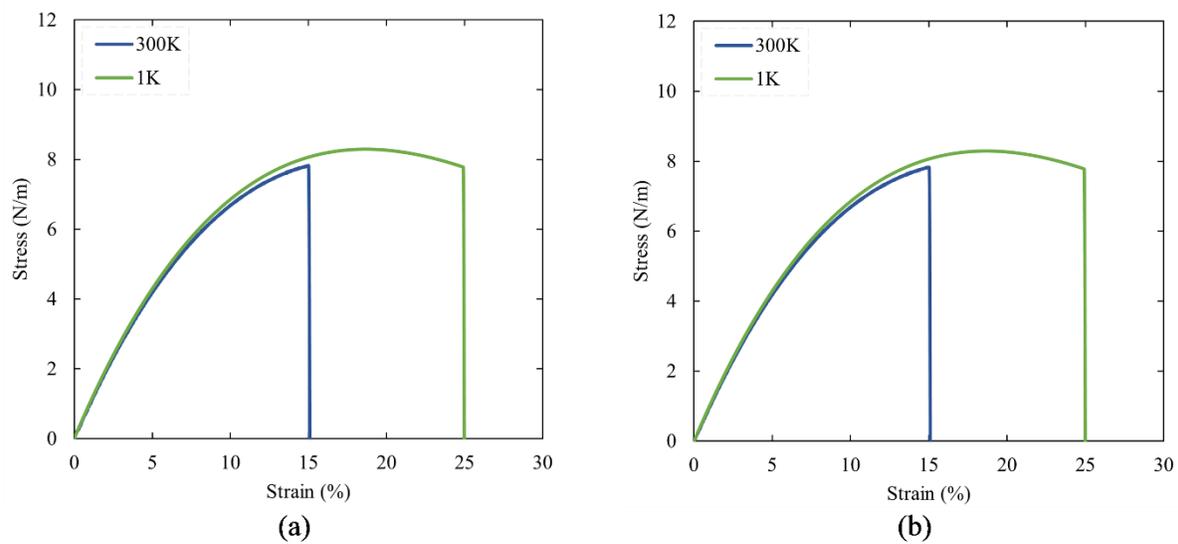

***Figure S 5*** *Tensile stress–strain curves of monolayer pristine 2H-MoTe$_2$ in (a) armchair direction and (b) zigzag direction under biaxial loading.*

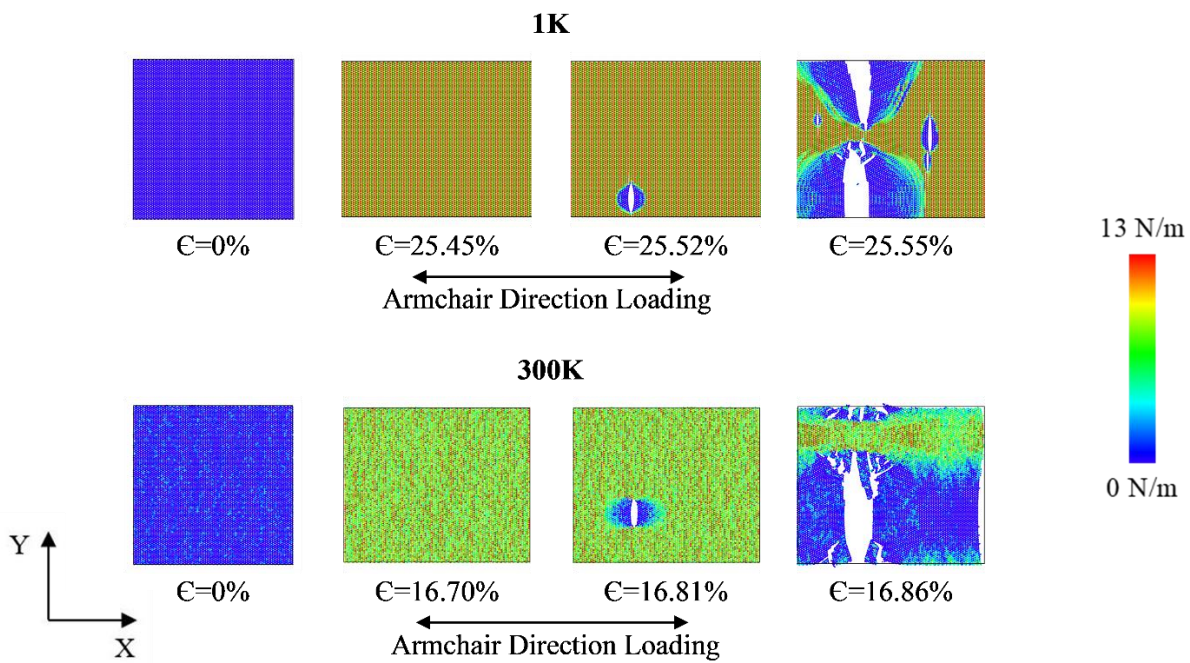

*Figure S 6* Crack propagation and stress distribution during armchair direction uniaxial tensile loading of monolayer pristine 2H-MoTe$_2$.

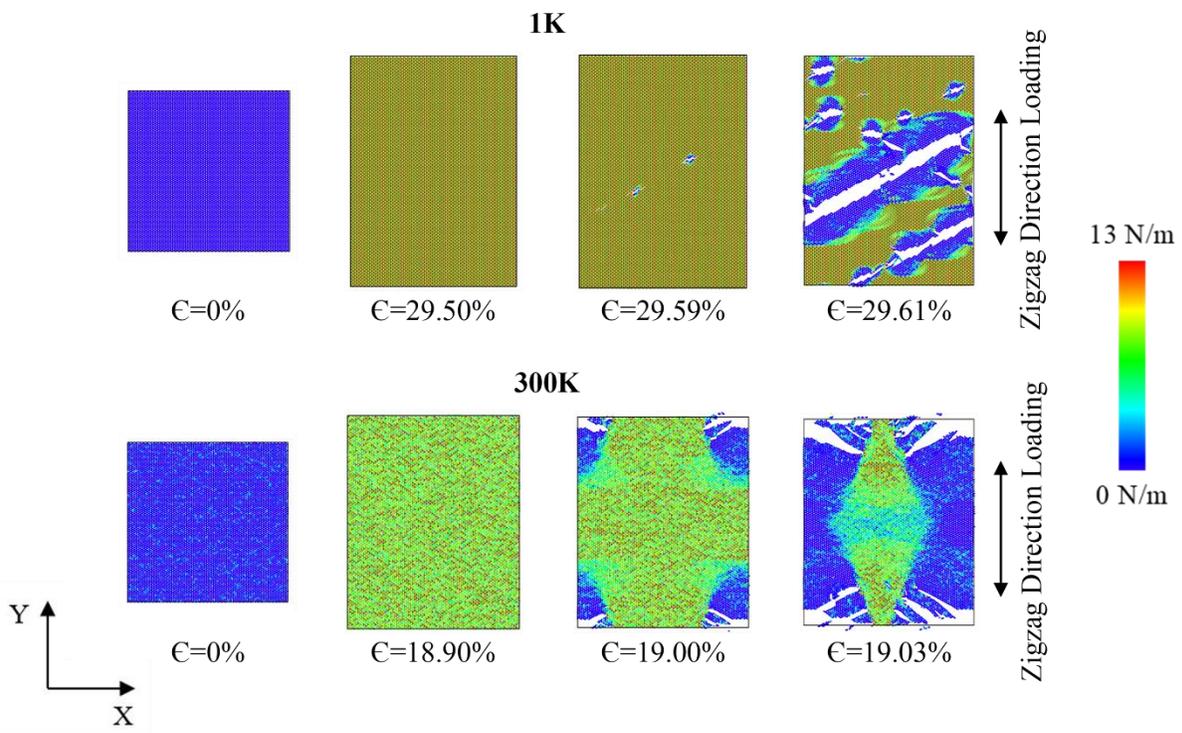

*Figure S 7* Crack propagation and stress distribution during zigzag direction uniaxial tensile loading of monolayer pristine 2H-MoTe$_2$.

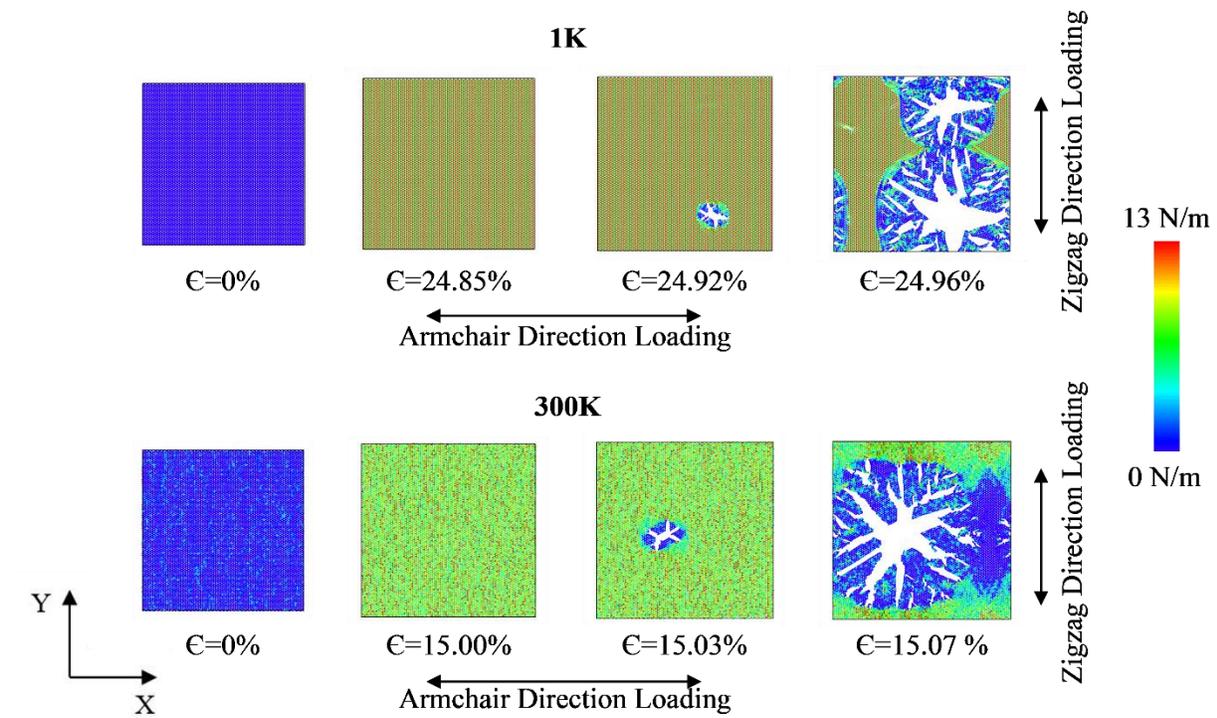

*Figure S 8* Crack propagation and stress distribution during biaxial tensile loading of monolayer pristine 2H-MoTe$_2$.